%% file: main.tex
\documentclass[conference]{IEEEtran}
\IEEEoverridecommandlockouts
\usepackage{cite}
\usepackage{amsmath,amssymb,amsfonts}
\usepackage{algorithmic}
\usepackage{graphicx}
\usepackage{textcomp}
\usepackage{xcolor}
\usepackage{algorithmic}
\usepackage{graphicx}
\usepackage{xcolor}
\usepackage{color}
\usepackage{mathrsfs}
\usepackage[ruled,vlined,linesnumbered,commentsnumbered]{algorithm2e}
\usepackage{amsmath}
\usepackage{amsthm}
\newtheorem*{definition1}{\textbf{\textit{Definition 1}}}
\newtheorem*{definition2}{\textbf{\textit{Definition 2}}}
\usepackage{multirow} 
\usepackage{multicol}
\usepackage{epsfig}
\usepackage{caption}
\usepackage{subfigure}
\usepackage{mathtools}
\usepackage{textcomp}
\usepackage{enumitem}
\usepackage{array}
\usepackage{booktabs}
\usepackage{makecell}
\usepackage{float}
\usepackage[labelfont=bf]{caption}
\usepackage{amsmath,amsfonts,amssymb}
\setlist[itemize]{leftmargin=*}

\newcommand{\tabincell}[2]{\begin{tabular}{@{}#1@{}}#2\end{tabular}}
\def\BibTeX{{\rm B\kern-.05em{\sc i\kern-.025em b}\kern-.08em
    T\kern-.1667em\lower.7ex\hbox{E}\kern-.125emX}}
    
\begin{document}
\title{On Disambiguating Authors: Collaboration Network Reconstruction in a Bottom-up Manner}

\author{\IEEEauthorblockN{Na Li~$^{1}$, Renyu Zhu~$^{1}$, Xiaoxu Zhou~$^{1}$, Xiangnan He~$^{2}$, Wenyuan Cai~$^{3}$, Ming Gao~$^{1,4}$\thanks{\IEEEauthorrefmark{1}Ming Gao is corresponding author.}, Aoying Zhou~$^{1}$}
\IEEEauthorblockA{$^{1}$\textit{School of Data Science and Engineering, East China Normal University, }
Shanghai, China\\
nali0606@foxmail.com, \{52175100003,51185100014\}@stu.ecnu.edu.cn, \{mgao,ayzhou\}@dase.ecnu.edu.cn\\
~$^{2}$\textit{School of Information Science and Technology, University of Science and Technology of China, }
Hefei, China\\
xiangnanhe@gmail.com\\
~$^{3}$\textit{Shanghai Hypers Data Technology Inc.,}
Shanghai, China, wenyuan.cai@hypers.com\\
~$^{4}$\textit{KLATASDS-MOE, School of Statistics, East China Normal University,}
Shanghai, China}
}

\maketitle

\begin{abstract}
    Author disambiguation arises when different authors share the same name, which is a critical task in digital libraries, such as DBLP, CiteULike, CiteSeerX, etc. 
    While the state-of-the-art methods have developed various paper embedding-based methods performing in a top-down manner, they primarily focus on the ego-network of a target name and overlook the low-quality collaborative relations existed in the ego-network.
    Thus, these methods can be suboptimal for disambiguating authors.
    
    In this paper, we model the author disambiguation as a collaboration network reconstruction problem, and propose an incremental and unsupervised author disambiguation method, namely IUAD, which performs in a bottom-up manner.
    Initially, we build a stable collaboration network based on stable collaborative relations.
    To further improve the recall, we build a probabilistic generative model to reconstruct the complete collaboration network.
    In addition, for newly published papers, we can incrementally judge who publish them via only computing the posterior probabilities.
    We have conducted extensive experiments on a large-scale DBLP dataset to evaluate IUAD.
    The experimental results demonstrate that IUAD not only achieves the promising performance, but also outperforms comparable baselines significantly. 
    Codes are available at https://github.com/papergitgit/IUAD.
\end{abstract}


\input{introduction}
\input{related_work}

\input{problem_definition}
\input{stable_network}

\input{global_network}
\input{discussion}

\input{experiments}

\input{conclusion}
\input{acknowledgement}

\input{mybib}


\end{document}

%% file: introduction.tex
\section{INTRODUCTION}\label{sec:introduction}
Author disambiguation, which aims at identifying the distinct authors shared with the same name from the paper database, is an important yet challenging task in many applications, especially in the online bibliography systems, such as DBLP, CiteULike, CiteSeerX, and so on. 
For example, searching for ``Wei Wang'' in DBLP returns 224 entries.
Not only they have the same name, but also many of them work in similar research fields. 
Although different authors have different emails and affiliations, these information is difficult to obtain if you do not read the papers.
It is thus a natural question that how to accurately disambiguate such authors.
For convenience, the word ``author'' represents a unique individual in this paper, while a name may be shared by multiple authors.
 
Author disambiguation is related to several similar tasks like record linkage~\cite{winkler1999state,winkler2006overview,fellegi1969theory}, entity resolution~\cite{bhattacharya2006latent,agarwal2018dianed,moon2018multimodal,nie2018mention}, object identification~\cite{tejada2001learning}, duplicate detection~\cite{bilenko2003adaptive,xu2017online,delgado2014data} and entity matching~\cite{shen2005constraint,li2019adversarial,zhang2016pct,liu2016aligning}, etc., being helpful to many applications in database, information retrieval, and data mining. 
To date, existing solutions can be roughly classified into two categories: supervised methods~\cite{han2004two, treeratpituk2009disambiguating, hermansson2013entity,kim2019hybrid,atarashi2017deep} and unsupervised methods~\cite{wang2008name,zhang2017name, xu2018network, tang2012unified, song2007efficient, bhattacharya2006latent, schulz2014exploiting, fan2011graph, shin2014author,zhang2019author,peng2019author}.
Since supervised learning methods require manual efforts to do data annotation and feature engineering, they are less transferable to new domains and are not suitable for large-scale applications. 

As such, recent advances in author disambiguation primarily focus on unsupervised methods~\cite{zhang2017name, xu2018network, mondal2019graph, shin2014author}.
Among them, embedding-based approaches achieve state-of-the-art performance~\cite{zhang2017name, xu2018network}.
Based on the ego-network of a target name, they embed all papers into low-dimensional vectors and then cluster them into clusters.
All papers of a cluster are considered as published by an identical author.
Despite effectiveness, we argue that such methods suffer from three limitations:

\begin{figure}[t]
  \centering
  \includegraphics[width=1\linewidth]{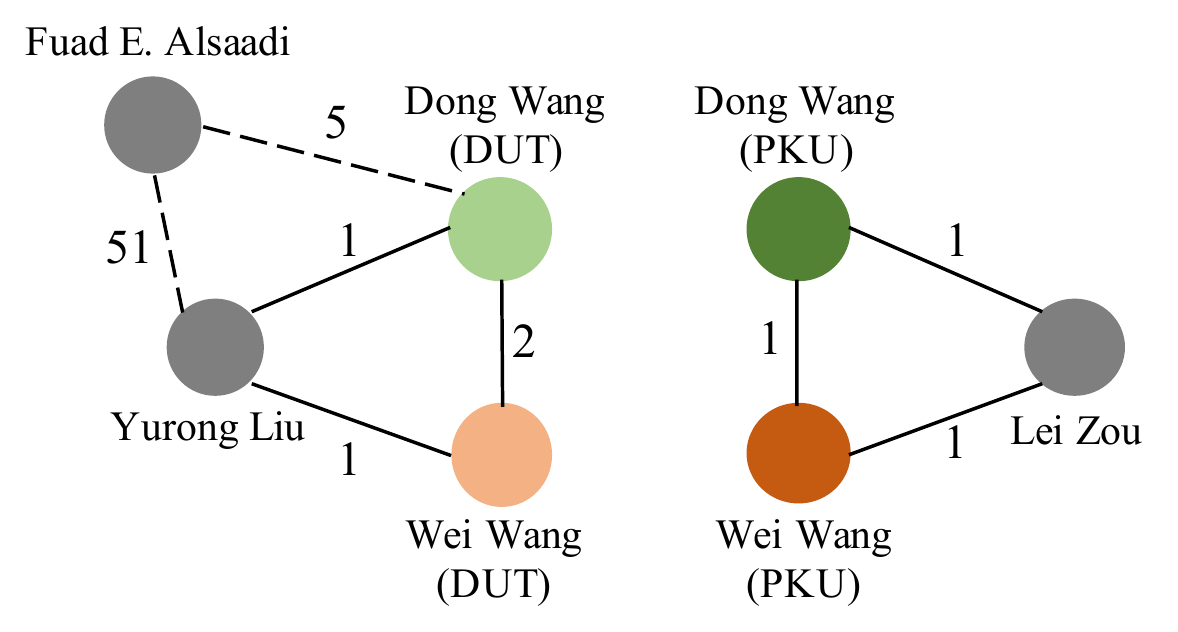}
  \caption{A running example of the collaboration network, where edges in dashed lines do not belong to the ego-network of ``Wei Wang (DUT)''.}
  \label{fig:unfrequent_overlap_cn}
  \vspace{-0.5cm}
\end{figure}

\begin{itemize}
\item They largely focus on clustering vertices in the ego-network, and perform in a top-down manner, i.e., all authors with the same name are represented by a vertex in the ego-network.
As such, the ego-network may not fully preserve the collaborative relations.
As demonstrated in Figure~\ref{fig:unfrequent_overlap_cn}, ``Dong Wang'', ``Yurong Liu'' and ``Wei Wang'' are likely to belong to the same community. However, the community structure is ignored by the ego-network of the target name ``Wei Wang''.


\item The ego-network may capture the inaccurate collaborative relations since all authors with the same name are treated as an identical author.
Actually, there are two authors named ``Dong Wang'', who collaborate with different authors named ``Wei Wang'' as illustrated in Figure~\ref{fig:unfrequent_overlap_cn}.
In this case, the wrong collaborative relations will be preserved in the ego-network.

\item They are mostly designed for dealing with static data.
For newly published papers, they require a full re-training to refresh the embedding vectors of papers. 
\end{itemize}

To address the limitations of the state-of-the-art approaches, we propose to perform author disambiguation by reconstructing the complete collaboration network in a bottom-up manner.
To the best of our knowledge, this is the first work to disambiguate authors in a bottom-up manner, which initially assumes that all authors with the same name are different.
Instead of mining low-quality collaborative relations, in the beginning, we extract the accurate collaborative relations from the co-author lists of papers.
To further balance precision and recall of our approach, we build a probabilistic generative model to further disambiguate authors via integrating diversified information, such as network structures, research interests, and research communities.
Once we learn the parameters of the probabilistic generative model, we can disambiguate authors for a newly published paper incrementally.
Thus, the main contributions of our proposed IUAD method are as followings:
\begin{itemize}
\item To address the limitations of existing unsupervised approaches, we propose an incremental and unsupervised author disambiguation method, namely IUAD, which consists of two stages.
To the best of our knowledge, IUAD is the first work for addressing author disambiguation in a bottom-up manner. 

\item In the first stage, we find the stable collaborative relations, which are defined by the frequent co-author relationships, from the co-author lists. 
Based on them, we build a stable collaboration network (SCN) to guarantee high precision of IUAD.

\item In the second stage, to further improve the recall, we adopt a general probabilistic generative model, which utilizes the exponential family to integrate diversified information, such as network structures, research interests, and research communities, to recover the global collaboration network (GCN) via merging vertices in the SCN.

\item For newly published papers, we can incrementally and efficiently judge who publish these papers in the global collaboration network. 

\item We conduct extensive experiments on the DBLP dataset to fully evaluate the performance and efficiency of IUAD.
The experimental results demonstrate that IUAD outperforms the state-of-the-art methods significantly.
\end{itemize}

%% file: related_work.tex
\section{RELATED WORK}\label{sec:related_work}

Our work is related to record linkage~\cite{fellegi1969theory, winkler1999state, winkler2006overview}, entity resolution~\cite{bhattacharya2006latent,agarwal2018dianed,moon2018multimodal,nie2018mention}, object identification~\cite{tejada2001learning}, duplicate detection~\cite{bilenko2003adaptive,xu2017online,delgado2014data} and entity matching~\cite{shen2005constraint,li2019adversarial,zhang2016pct,liu2016aligning}, etc., which are applied in many scenes widely, such as social network linkage, data integration, database de-duplication, and so on.
Existing approaches can be classified into two categories: supervised and unsupervised methods. 

\subsection{Supervised Methods}
Some existing works train classifiers to address the author disambiguation problem~\cite{han2004two, amancio2015topological, treeratpituk2009disambiguating, hermansson2013entity,kim2019hybrid,atarashi2017deep}.
Han et al. adopt coauthor names, paper titles, and journal titles, etc., to train classifiers to disambiguate authors~\cite{han2004two}. 
Treeratpituk et al. extract a set of features, including similarities of authors, affiliations, coauthors, concepts, journals and titles, etc., to address the author disambiguation problem in scientific databases~\cite{treeratpituk2009disambiguating}. 

For supervised approaches, they need a lot of labeled data to train models, which require labor overhead for data annotation.
We, therefore, propose an unsupervised approach to address the author disambiguation problem.

\subsection{Unsupervised Methods}
Since unsupervised methods do not need to collect the labeled data, they are suitable to apply for addressing large-scale author disambiguation problems.
As such, recent advances in author disambiguation have primarily focused on unsupervised methods~\cite{zhang2017name, xu2018network, schulz2014exploiting, shin2014author, mondal2019graph,zhang2018name,zhang2019author,peng2019author}.

However, most of them are top-down approaches.
When these top-down approaches construct the ego-networks, they treat all authors shared the same name as an identical author~\cite{bhattacharya2006latent, shin2014author,tang2012unified, mondal2019graph, liu2015fast, schulz2014exploiting, zhang2017name, zhang2018name,zhang2019author,peng2019author}.
Zhang et al. model the author disambiguation as a clustering problem after embedding each paper into a low-dimensional space~\cite{zhang2017name}.
Similarly, Xu et al. employ five types of networks to embed papers into a low-dimensional space, and further cluster papers into groups~\cite{xu2018network}.
Peng et al. adopt Generative Adversarial Networks to learn the paper representation of the heterogeneous network, then, HDBScan and AP are used to cluster papers~\cite{peng2019author}.
Shin et al. tackle this problem by splitting vertices in the graph of co-authorships~\cite{shin2014author}.
Liu et al. introduce a coarse-to-fine multiple clustering framework~\cite{liu2015fast}.
Despite effectiveness, we argue that such methods suffer from the following limitations: (1) the top-down approaches tend to initially mine the low-quality collaborative relations since they do not distinguish authors in the ego-networks; (2) they are designed for dealing with the static data, cannot handle the newly published papers incrementally. 

Fan et al. are conscious of the limitations of existing unsupervised methods~\cite{fan2011graph}.
For multiple papers published by a target name, they do not merge the authors with the target name into a single vertex in the ego-network.
However, they do not distinguish the names of their co-authors, i.e., two different co-authors sharing the same name are treated as a unique co-author.
In addition, they only utilize the network structure to disambiguate authors.
Actually, the other information, such as paper titles, published venues, and research communities, is helpful to disambiguate authors.

To the best of our knowledge, our proposed IUAD approach is the first work to perform author disambiguation in a bottom-up manner.
In the beginning, instead of mining the low-quality collaborative relations, we extract the stable collaborative relations from the co-author lists.
To further improve the performance of recall, we build a probabilistic generative model to further disambiguate authors via integrating diversified information, such as paper titles, published venues, and research communities.

%% file: problem_definition.tex
\section{PROBLEM FORMULATION}
\label{sec:problem_definition}

\subsection{Problem Definition}
\label{subsec:problem_definition}

In this paper, we disambiguate authors via reconstructing the collaboration network.
Our input is a paper database $\mathcal{D}$, where each paper has four attributes: co-author list, title, published venue, and published year. 
Our goal is to reconstruct the collaboration network, which is defined as follows:
\begin{definition1}{\textit{\textbf{Collaboration Network}}}
\label{def:cn}
Given a paper database $\mathcal{D}$ associated with author set $\mathcal{V}$, a collaboration network is a graph $G = (V, E, P)$, where $V$ is a vertex set, such that $E \subset V \times V$, and edge $(u,v)\in E$ associates with a set of papers, denoted as $P^{uv}$, and all papers in  $P^{uv}$ are published by coauthors $u$ and $v$.
\end{definition1}

Based on the above definition, the author disambiguation task can be treated as a collaboration network reconstruction problem. 
We propose a two-stage, incremental, and unsupervised author disambiguation algorithm, namely IUAD, and reconstruct the collaboration network in a bottom-up manner. 

\subsection{General Framework}
\label{subsec:framework}
\begin{figure}[t]
  \centering
  \includegraphics[width=1\linewidth]{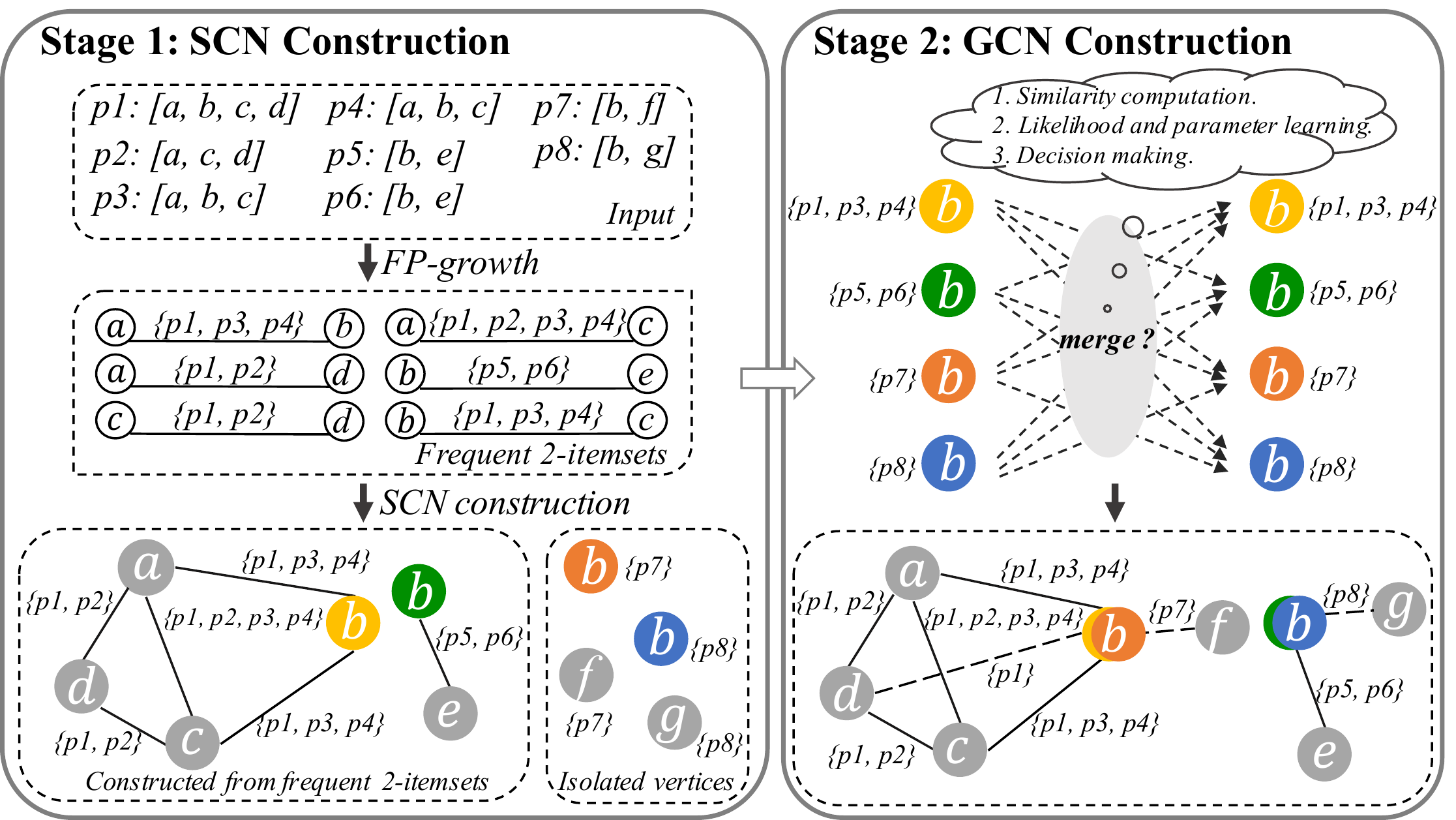}
  \caption{\textbf{The framework of IUAD.
  }}
  \label{fig:framework}
    \vspace{-0.7cm}
\end{figure}

In our solution, we propose a two-stage author disambiguation algorithm IUAD.
In the first stage, we build a stable collaboration network (SCN) to capture accurate and higher-order collaborative relations.
In the second stage, we construct a global collaboration network (GCN) via applying a probabilistic generative model to judge whether two vertices in the stable collaboration network belong to a unique author or not.
The framework of IUAD is illustrated in Figure~\ref{fig:framework}.

As demonstrated in Figure~\ref{fig:framework} and Algorithm~\ref{alg:IUAD}, IUAD consists of two sub-tasks: stable collaboration network (SCN) construction and global collaboration network (GCN) construction.
In the first stage (Lines 2-5), IUAD forms the stable collaboration network via mining all $\eta-$stable collaborative relations and their formed triangles (Line 5).
In the second stage (Lines 7-15), IUAD further merges vertices of SCN via adopting a probabilistic generative model (Line 9), where a vertex in SCN can be modeled as a subset of papers published by author $a$.
IUAD employs the EM algorithm to solve the model at Line 10. For each name $a\in \mathcal{A}$, IUAD computes a score for every vertex pair $(v_i^a, v_j^a)$ to judge whether two vertices $v_i^a$ and $v_j^a$ belong to an identical author or not at Lines 13-15.
Finally, IUAD recovers the collaborative relations existing in the paper co-author lists (Line 16).

IUAD performs a good trade-off between precision and recall.
In the first stage, SCN mines stable collaborative relations to capture the higher-order and higher-quality collaborative information, which ensures the high precision of IUAD.
In the second stage, GCN captures the identical authors in the SCN as many as possible, which is responsible for improving the recall of IUAD.  

\IncMargin{1em} 
\begin{algorithm}
    \SetKwInOut{Input}{\textbf{Input}}\SetKwInOut{Output}{\textbf{Output}} 
    \Input{
    A set of papers $\mathcal{D}$ with name set $\mathcal{A}$, and decision threshold $\delta$.}
    \Output{
    Global collaboration network $G$.
    }
    \BlankLine
    \emph{//\textbf{Stage I: SCN Construction}}\; 
    \For{$\forall p \in \mathcal{D}$}{
        $L_p\leftarrow$ co-author list of paper $p$\; 
    }
    Generate all $\eta-$SCRs $\mathcal{F}$ from $\{L_p|p\in \mathcal{D}\}$\;
    Construct SCN $G$ based on $\mathcal{F}$ and the remaining papers\;
    \emph{//\textbf{Stage II: GCN Construction}}\;
    \For{two vertices in $G$ with the identical name}{
       $\Gamma \leftarrow$ compute the similarity vector between them\; 
    }
    Build a probabilistic generative model\;
    Learn parameter vector $\hat{\Theta}$ by using the EM algorithm\;
     \For{each $a\in \mathcal{A}$}{
         \For {each vertex pair $r = (v_i^a, v_j^a)$ with similarity vector $\gamma\in \Gamma$}{
            $s \leftarrow log(\frac{P(r\in M |\gamma, \hat{\Theta})}{P(r\in U |\gamma, \hat{\Theta})})$, where $M$ and $U$ are the sets of matched and unmatched pairs, respectively\;
            \If{$s \geq \delta$}{
            Merge vertices $v_i^a$ and $v_j^a$ as a single vertex\;
            }
        }
    }
   Recover collaborative relations existed in $L_p$\;
   \textbf{return} Global collaboration network $G$\;
    \caption{\textbf{IUAD} algorithm\label{alg:IUAD}}
\end{algorithm}
\DecMargin{1em}

%% file: stable_network.tex
\section{Stable Collaboration Network Construction}
\label{sec:stable_network}
Instead of mining the low-quality collaborative relations in the existing methods, we design a bottom-up method, which aims at finding the stable collaborative relations from the co-author lists, and further construct the stable collaboration network.

\subsection{Key Observation}
\label{subsec:observation}
In a collaboration network, a name exists in the co-author list can be considered as a random event.
Let $n_a$ and $n_b$ be the numbers of papers published by names $a$ and $b$, respectively, and $N$ be the total number of papers in paper database $\mathcal{D}$.
Assume that $a$ and $b$ independently appears in the co-author list of paper $p\in \mathcal{D}$, we have $Pr(a\in p, b\in p) = \frac{n_a}{N}\cdot \frac{n_b}{N}$ (we will infer that the independent assumption is incorrect). Let $X_i$ be a Bernoulli r.v., such that,
\[X_i = \left\{
  \begin{array}{ll}
    1, & \hbox{names } a \hbox{ and } b \hbox{ exist in the co-author list of }p_i;  \\
    0, & \hbox{otherwise.}
  \end{array}
\right.\]
Thus, $Pr(X_i = 1) = \frac{n_an_b}{N^2}$, and $X = \sum_{i=1}^{N}X_i$ is the number of papers co-authored by $a$ and $b$.
We know that $X \sim Binom(N, \frac{n_an_b}{N^2})$. 
Under the independent assumption, we approximate the probability of names $a$ and $b$ co-exist in papers at least $x$ times, i.e., $Pr(X \geq x)$.
According to the weakly Central Limit Theorem, we can approximate the probability by using the standard normal distribution.
The probability can be computed as:
\begin{equation}
    \begin{aligned}
        Pr(X \geq x) & = 1 - Pr(X < x - 0.5)  \\
        &\approx 1 - \Phi(\frac{(x - 0.5)-E(X)}{\sqrt{var(X)}}),
    \end{aligned}
\end{equation}
where ``$-0.5$" is to convert discrete case to continuous case, the standardized r.v. $\frac{X-E(X)}{\sqrt{var(X)}}$ can be approximated as $\mathcal{N}(0,1)$, and $\Phi(\cdot)$ is a CDF of $\mathcal{N}(0,1)$. 
\begin{figure}[tbp]
      \centering
      \subfigure[\textbf{\# papers per name}]{\includegraphics[width=0.445\linewidth]{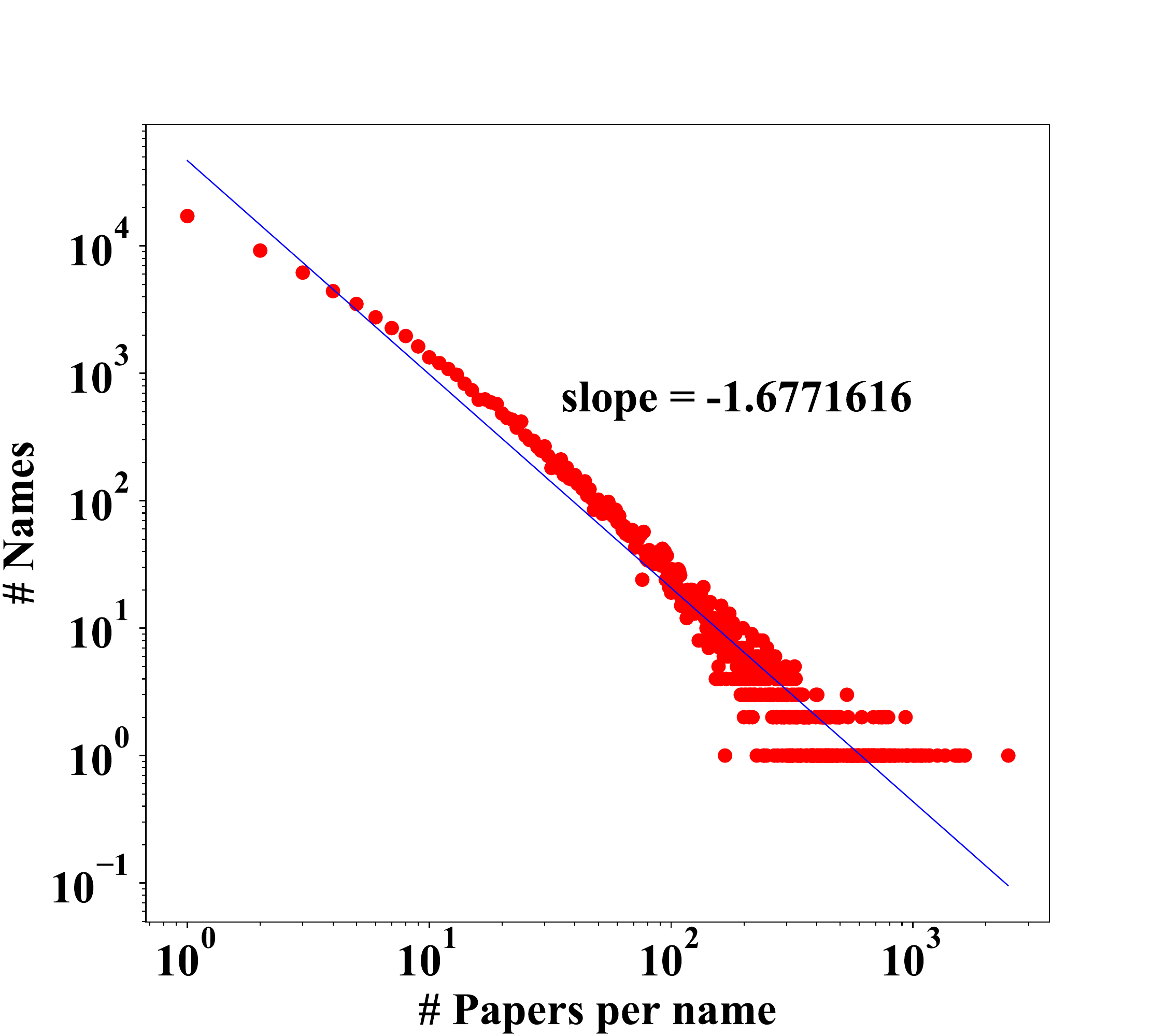}}
      \subfigure[\textbf{\# frequent itemsets}]{\includegraphics[width=0.445\linewidth]{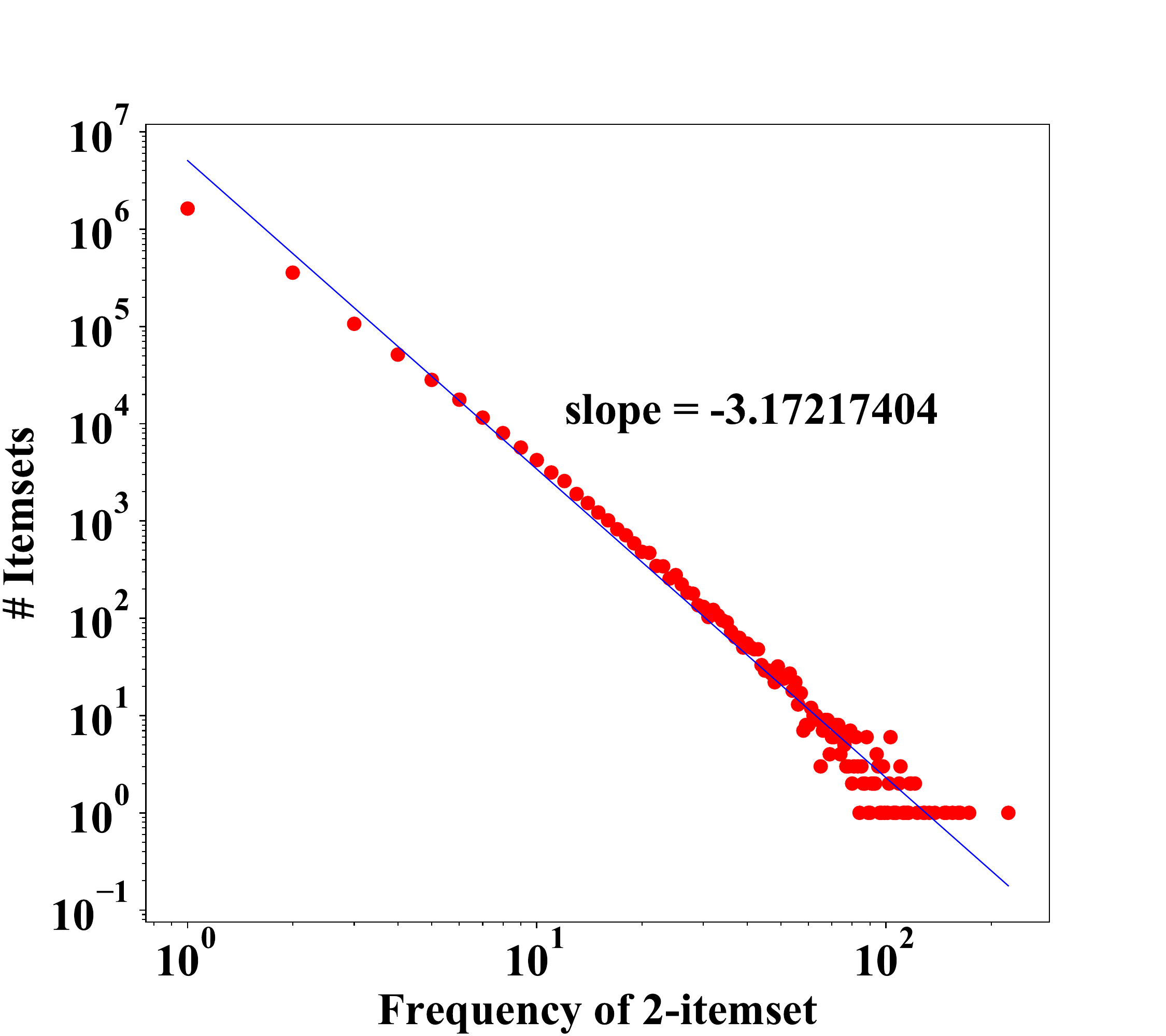}}
      \caption{Descriptive analysis of DBLP dataset.}
      \label{distribution}
    \end{figure}

As illustrated in Figure~\ref{distribution}(a), we can observe that the average number of papers, which are published by the same name, is less than 500. Suppose $n_a = 5\times 10^2, n_b= 5\times 10^2, N=5\times 10^5$, then $E(X) = N \cdot \frac{n_a n_b}{N^2} = 0.5$, $Var(X) =N \cdot \frac{n_a n_b}{N^2} \cdot (1- \frac{n_a n_b}{N^2})\approx 0.5$. Let $x=3$, the probability of $P(X \geq x)$ is:
\begin{equation}
    \begin{aligned}
    Pr(X \geq 3) &=1 - \Phi( \frac{2.5-E(X)}{\sqrt{var(X)}}) = 2.3389 \times 10^{-3}.
    \end{aligned}
\end{equation}
From this case, the tail probability $Pr(X \geq 3)$ is very small, i.e., the probability of names $a$ and $b$ co-exist in papers at least $3$ times is very small, and the tail probability will become smaller if the values of $n_a$ and $n_b$ decrease. 

Surprisingly, we can observe that the frequencies of name pairs follow the power-law distribution in co-author lists as illustrated in Figure~\ref{distribution}(b).
This indicates that an author tends to collaborate with others who collaborate with him/her frequently.
This phenomenon proves that the independence assumption for name co-occurrence is incorrect.
That is, two authors $a_1$ and $a_2$ share name $a$, and the other two authors $b_1$ and $b_2$ also share name $b$, then it is almost an impossible event that both author pairs $(a_1,b_1)$ and $(a_2,b_2)$ have the high frequencies in the co-author lists.

Thus, we have strong evidence to conclude that all papers co-authored by $a$ and $b$ are published by a unique author $a$ or a unique author $b$ if name pair $(a,b)$ frequently appears in the co-author lists.
This is due to the fact that a collaborative relationship, which is formed in a scale-free network, is not a random event~\cite{erdos1959random}.

\subsection{Stable Collaborative Relation}
\label{subsec:scn}
Based on the above observations, we define the following stable collaborative relation as follows:
\begin{definition2}{\textit{\textbf{$\eta$-Stable Collaborative Relation}}}
Given a paper database $\mathcal{D}$ associated with name set $\mathcal{A}$, name pair $(a,b)$ forms an $\eta-$stable collaborative relation (short in \textbf{SCR}) if co-occurrence frequency of names $a$ and $b$ is no less than $\eta$ in all co-author lists for $a,b\in \mathcal{A}$.
\end{definition2}
Note that the $\eta-$SCR is a symmetric relation.
According to the definition, we can find all $\eta-$SCRs via mining all frequent itemsets with support threshold $\eta$ from the co-author lists of papers.

Once we find all $\eta-$SCRs, we will accurately recover a large number of collaborative relations in the collaboration network.
Furthermore, a triangle, which is formed in a scale-free network, is also not a random event.
This is because that the number of triangles, which a vertex participates in, also follows the power-law pattern in a scale-free network~\cite{Tsourakakis08}.
Thus, if $(a,b)$, $(a,c)$ and $(b,c)$ are $\eta-$SCRs, then $(a,b,c)$ also forms a stable collaborative triangle.
Thus, we construct a \textbf{stable collaboration network}, short in \textbf{SCN}, which preserves all $\eta-$SCRs or their formed triangles.

\subsection{Stable Collaboration Network}
\label{subsec:scn_construction}
Next, we will address how to construct the stable collaboration network from the input paper database. 
The construction consists of two steps:
\begin{itemize}
\item \textbf{Step I: Generating $\eta-$Stable Collaborative Relations.} 
We employ the \textit{FP-growth} algorithm~\cite{han2000mining} with support threshold $\eta$ to mine all $\eta-$SCRs, denoted as $\mathcal{F}$, from the co-author lists.
If name pair $(a,b)\in \mathcal{F}$, $a$ and $b$ will be two vertices in the SCN, and an edge will be formed between them.
In the SCN, a vertex is treated as a unique author, i.e., all papers co-authored by $a$ and $b$ are published by a unique author $a$ or unique author $b$ if $(a,b)$ forms a $\eta-$SCR.
Thus, edge $(a,b)$ in the SCN associates with a set of papers, denoted as $P^{ab}$, which are co-authored by authors $a$ and $b$.

\item \textbf{Step II: Constructing SCN.} 
To capture higher-order collaborative relations, we further infer the stable triangles from the found $\eta-$SCRs. 
Let $(a,b)$, $(a,c)$, and $(b,c)$ be $\eta-$SCRs, then $(a,b,c)$ will form a stable collaborative triangle, i.e., the papers, where their co-author lists contain one of $\eta-$SCRs $(a,b)$, $(a,c)$, and $(b,c)$, are from the identical authors $a$, $b$ or $c$. 
\end{itemize}

\begin{figure}[t]
      \centering
      \includegraphics[width=0.9\linewidth]{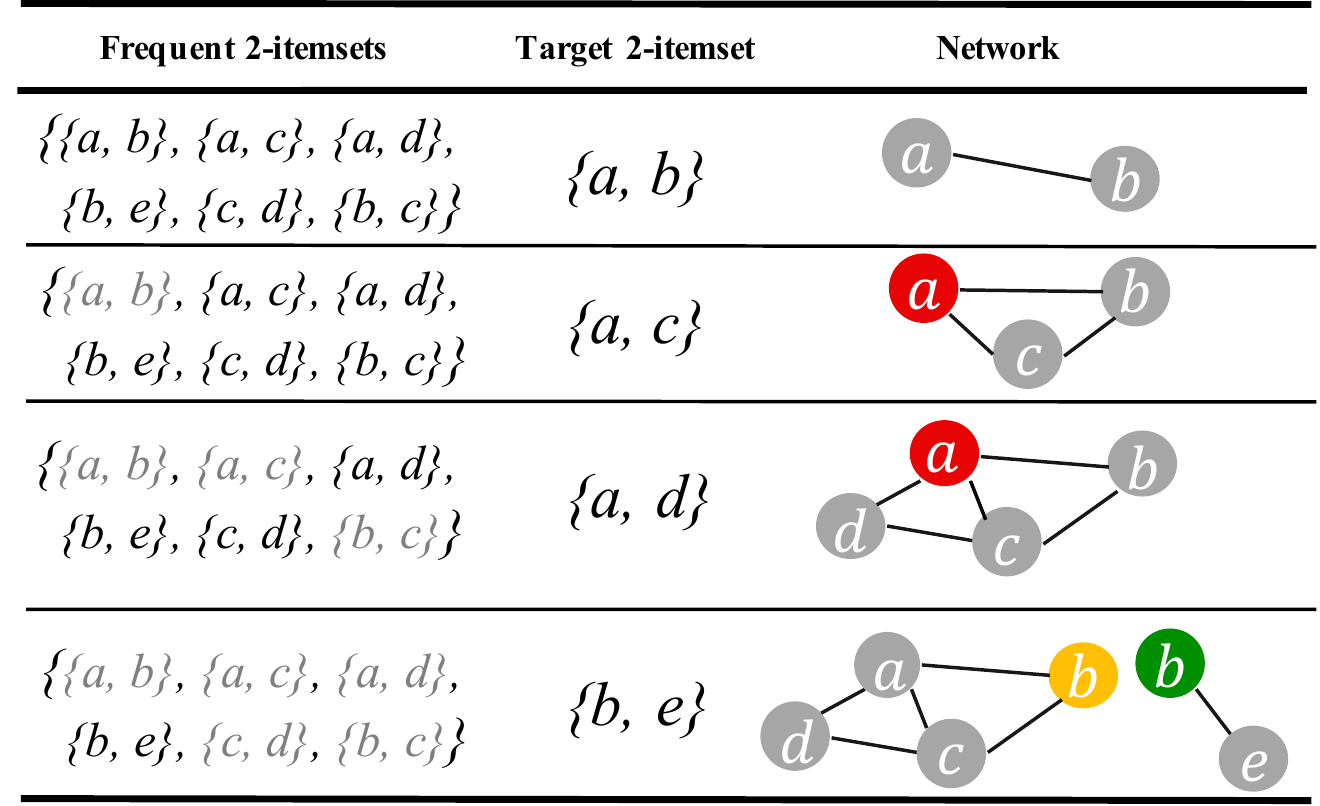}
      \caption{\textbf{A running example of SCN construction (frequent 2-itemsets construction part.)}
      }
      \label{fig:generate}
\end{figure}

To better understand the SCN construction, Figure~\ref{fig:framework} demonstrates a running example and Figure~\ref{fig:generate} illustrates the detailed construction process, where $(a,b)$, $(a,c)$, $(a,d)$, $(b,e)$, $(c,d)$, and $(b,c)$ are 2-SCRs.
\begin{enumerate}
    \item[(i)] Insert 2-SCR $(a,b)$: Before inserting it, SCN $G$ is an empty network. Vertices $a$ and $b$ will insert into network $G$, and an edge $(a,b)$ is also formed.
    
    \item[(ii)] Insert 2-SCR $(a,c)$: Since $a$ is a vertex of $G$, we need to judge whether authors $a$ in 2-SCRs $(a,b)$ and $(a,c)$ are identical or not. Searching the set of 2-SCRs, we find that $(b,c)$ is also a 2-SCR. Thus, we consider authors $a$ in 2-SCRs $(a,b)$ and $(a,c)$ are identical. Furthermore, vertex $c$ is added into network $G$, and edges $(a,c)$ and $(b,c)$ are formed.
    
    \item[(iii)] Insert 2-SCR $(a,d)$: Similar to 2-SCR $(a,c)$, vertex $d$ is added into network $G$, and edges $(a,d)$ and $(c,d)$ are formed after inserting 2-SCR $(a,d)$.
    
    \item[(iv)] Insert 2-SCR $(b,e)$: Although $b$ is a vertex of $G$, there is not a 2-SCR between names $e$ and existing neighbor vertices of $b$. Therefore, vertices $b$ and $e$ will be inserted into the network as new vertices, and edge $(b,e)$ is also formed.
    
    \item[(v)] Insert remaining names: The authors, who do not exist in any 2-SCRs, will be inserted into the network as isolated vertices.
\end{enumerate}

In this stage, we construct the SCN which captures the stable and higher-order collaborative relations from the paper database. 
As we analyzed, all edges of a SCN are stable collaborative relations.
Thus, the constructed SCN ensures the high precision of IUAD.

%% file: global_network.tex
\section{Global Collaboration Network Construction}
\label{sec:global_network}
In the SCN construction stage, we ensure the high precision of IUAD by mining $\eta$-SCRs.
However, due to the changes in research interests, the collaboration network may change over time.
Therefore, two vertices with the same name in SCN may be the identical author, i.e., multiple vertices of SCN may belong to an identical author.
To improve the recall, IUAD further judges whether two vertices with the same name are a unique author or not. 
Finally, we reconstruct a \textbf{global collaboration network}, short in \textbf{GCN}.

\subsection{Problem Formulation}
\label{subsec:problem_formulation}

Let $V^a=\{v_1^a, v_2^a, \cdots, v_n^a\}$ be the set of vertices with name $a$ in SCN, $R^a \subset V^a \times V^a = \{(v_i^a, v_j^a)| v_i^a \in V^a, v_j^a \in V^a, i \neq j \}$ be the set of candidate pairs, and $R = \bigcup_{a\in \mathcal{A}}R^a$.
For $r_j\in R$, $\gamma_j$ represents the similarity vector between two vertices in pair $r_j$.
The task of constructing GCN aims at determining whether two vertices are a unique author or not, i.e., $R = M\cup U$, where $M$ is the set of vertex pairs whose two vertices belong to a unique author (called the set of matched pairs), otherwise $U$ (called the set of unmatched pairs).

To solve the problem, we propose a generative probabilistic model to calculate the probabilities $Pr(r_j\in M|\gamma_j, \Theta)$ and $Pr(r_j\in U|\gamma_j, \Theta)$ to make the decision~\cite{gao2015cnl}, where $\Theta$ denotes the parameters of the generative model.
To calculate the probability, we need to compute the similarity vectors of candidate vertex pairs.

\subsection{Similarity Computation}
\label{subsec:similiarity_functions}

As we know, the topological structures of the collaboration network, research interests, and research communities are helpful to identify authors from the SCN.
We propose six similarity functions for vertex pair $r_j$, denoted as $\gamma_j = (\gamma^{(1)}_j, \gamma^{(2)}_j,\cdots, \gamma^{(6)}_j)$, to capture how similar between two vertices in SCN.

\subsubsection{\textbf{Similarities in Network Structures}} 
The network structures reflect the collaborative relations of authors, which are important to identify whether two vertices are a unique author or not.
Two similarity functions are defined in this part to measure similarities in topological structures of the collaboration network.

\textbf{Normalized Weisfeiler-Lehman Sub-graph Kernel.} 
The topological structures of the collaboration network can well reflect the similarity between vertices. Usually, paths, cycles, or kernels \cite{shin2014author, fan2011graph, hermansson2013entity} are used to measure the similarity of vertices.
Due to the inefficiency of computing paths and cycles with larger networks, IUAD adopts Weisfeiler-Lehman sub-graph kernel~\cite{shervashidze2011weisfeiler} to evaluate the similarity between two vertices.
WL-kernel captures topological information to quantify the similarity of vertices in SCN, which can judge how similar two vertices are.
The WL sub-graph kernel of two vertices $v_i^a$ and $v_j^a$ is defined as:
\begin{equation}
    \begin{aligned}
    K^{	\left \langle h \right \rangle}(v_i^a, v_j^a) = 	\left \langle \phi^{	\left \langle h \right \rangle}(v_i^a), \phi^{\left \langle h	\right \rangle}(v_j^a) 	\right \rangle,
    \end{aligned}
\end{equation}
where $K^{	\left \langle h	\right \rangle}(v_i^a, v_j^a)$ measures the similarity between vertices $v_i^a$ and $v_j^a$ based on the number of occurrences of co-authors in $h$-th iteration, which is denoted as $\phi^{	\left \langle h	\right \rangle}(\cdot)$.
To avoid the negative effect of different sub-graph sizes, the normalized WL sub-graph kernel is adopted: 
\begin{equation}
 \begin{aligned}
 \gamma^{(1)}_j = \frac{ K^{\left \langle h \right \rangle}(v_i^a, v_j^a)}{\sqrt{K^{\left \langle h \right \rangle}(v_i^a, v_i^a) \cdot K^{\left \langle h	\right \rangle}(v_j^a, v_j^a)}}.
 \end{aligned}
\end{equation}
Due to page limitation, more details please refer to ~\cite{shervashidze2011weisfeiler, ah2010normalized}.
For the WL sub-graph kernel, the more similar in topological structures of $v_i^a$ and $v_j^a$ are, the larger value of the WL sub-graph kernel is, and the more likely that two vertices are a unique author.

\textbf{Co-author Clique Coincidence Ratio.}
As analyzed before, triangles describe one kind of stable collaborative relations.
In general, a clique will form more stable collaborative relations.
In SCN, two vertices are more likely to be a unique author if they have many common cliques.
Co-author clique coincidence ratio measures how many co-author common cliques they have.
Let $L(v_i^a)$ and $L(v_j^a)$ are the sets of co-author cliques of vertices $v_i^a$ and $v_j^a$, respectively.
The co-author clique coincidence ratio is defined as:
\begin{equation}
    \begin{aligned}
    \gamma^{(2)}_j = \frac{1}{\tau} |L(v_i^a) \cap  L(v_j^a)|,
    \end{aligned}
\end{equation}
where $\tau$, which balances the productivity of authors, is the smaller number of papers published by $v_i^a$ and $v_j^a$.
To speed up the clique computation, we only list the triangles in $L(\cdot)$.
For this similarity, the higher the coincident degree of co-author clique is, the higher probability to be a unique author is, which is consistent with our intuition.

\subsubsection{\textbf{Similarity in Research Interests}} 
Different authors may have different interests, and the same author may have coherent interests over time.
Since paper titles reflect author interests, we define two similarity functions to measure similarities of author interests based on paper titles.

\textbf{Similarity of Research Interests.}
The keywords in paper titles reflect author interests.
Since the distributed representations of words preserve the semantic information of text,
we adopt word vectors learned by language models, such as Word2Vec, GloVe and BERT, etc., to capture the semantic of keywords.
The similarity of research interests can be measured by cosine similarity of word vectors, which is defined as follows:
\begin{equation}
    \begin{aligned}
    \gamma^{(3)}_j = \frac{ W(v_i^a) \cdot W(v_j^a)}{||W(v_i^a)||_2 \cdot ||W(v_j^a)||_2},
    \end{aligned}
\end{equation}
where $W(v_i^a)$ and $W(v_j^a)$ are the centers of all keyword vectors of vertices $v_i^a$ and $v_j^a$.
To extract the keywords, the stop words or the frequent words in paper titles are excluded.
Thus, the cosine similarity is larger, i.e., $v_i^a$ and $v_j^a$ are more similar in the research interests, the more probability that two vertices belong to a unique author.

\textbf{Time Consistency in Research Interests.}
For an author, although the research field may not change, the research problems vary over time.
Thus, we define a similarity to capture the time consistency in research interests.
Let multisets $B(v_i^a)$ and $B(v_j^a)$ are keywords that exist in paper titles of vertices $v_i^a$ and $v_j^a$, respectively.
The similarity of time consistency in research interests is defined as:
\begin{equation}
    \begin{aligned}
    \gamma^{(4)}_j = \frac{1}{\tau} \sum_{b \in B(v_i^a) \cap B(v_j^a)} e^{\alpha* \min{(b)}} * \frac{1}{log(FB(b))},
    \end{aligned}
\end{equation}
where $\alpha$ is a decay factor introduced in~\cite{sayyadi2009futurerank} (set to $0.62$ in our experiment), $\min{(b)}$ is the minimum year difference of word $b$, $FB(b)$ is occurrence frequency of word $b$ in all titles of the whole dataset, and $\tau$ is the same to the definition in $\gamma^{(2)}_j$.
From this definition, if more unfrequent words are used by authors $v_i^a$ and $v_j^a$, they have higher time consistency in research interests.

\subsubsection{\textbf{Similarity of Research Communities}}
Due to the cognitive limitations of the human~\cite{Dunbar1992}, authors may have stable research communities.
In this paper, we employ the venues of published papers to capture the similarity of the research communities. 

\textbf{Similarity of Representative Community.}
An author has a stable research field, which forms a research community, he/she may publish papers in the same venue frequently.
Let multiset $H(v)$ be the set of venues, in which papers published by vertex $v$, and $h_i^a$ and $h_j^a$ be the most frequent venues in $H(v_i^a)$ and $H(v_j^a)$, respectively.
Note that $h_i^a$ and $h_j^a$ are considered as the representative communities of vertices $v_i^a$ and $v_j^a$, respectively.
The following similarity function is defined to measure the similarity of representative community:
\begin{equation}
    \begin{aligned}
    \gamma^{(5)}_j &= \frac{1}{\tau}(cnt(H(v_j^a), h_i^a) + cnt(H(v_i^a), h_j^a)),
    \end{aligned} 
\end{equation}
where $cnt(H, h)$ denotes the frequency of venue $h$ existed in multiset $H$, and $\tau$ is the same as the definition in $\gamma^{(2)}_j$.
If vertices $v_i^a$ and $v_j^a$ frequently publish papers in the same venue, they have high representative community similarity.

\textbf{Similarity of Research Communities.} 
Apart from the representative community, all venues are helpful to disambiguate authors.
Especially, some authors may have special research communities, which can be represented as small minority venues.
Referring to the Adamic/Adar metric, we define the following similarity function to measure the similarity of the research communities:
\begin{equation}
    \begin{aligned}
    \gamma^{(6)}_j= \frac{1}{\tau} \sum_{h \in H(v_i^a) \cap H(v_j^a)} \frac{1}{log(FH(h))},
    \end{aligned}
\end{equation}
where $FH(h)$ is the number of papers published in venue $h$ among the entire dataset, and $\tau$ is the same to $\gamma^{(2)}_j$.
In this similarity function, it emphasizes the small minority venues to disambiguate authors.
If two vertices $v_i^a$ and $v_j^a$ frequently publish papers in small minority venues, they are more likely to be a unique author.

\subsection{Likelihood and Parameter Learning}
\label{subsec:parameter_learning}

For a candidate vertex pair $r_j$, $P(r_j \in M|\Theta)$ and $P(r_j \in U|\Theta)$ represent the matched and unmatched probabilities, respectively.
If we suppose that $P(r_j \in M|\Theta) = p$, then $P(r_j \in U|\Theta) = 1 - p$.
Let $l_j = 1$ if $r_j \in M$, otherwise $l_j = 0$, and binary vector $\mathbf{l}^j = (l_j, 1 - l_j)$.
We define $x_j=(\mathbf{l}^j, \gamma_j) $ as the ``complete data'' for $r_j$. The log-likelihood is:

\begin{equation}
    \begin{aligned}
    \mathbb{L}(\Theta|X) &=
    \sum_{j=1}^{N}\mathbf{l}^j\Big[\log{P(\gamma_j|r_j\in M,\Theta)},\log{P(\gamma_j|r_j\in U,\Theta)}\Big]^T \\
&+\sum_{j=1}^{N}\mathbf{l}^j\Big[\log{p},\log{(1-p)}\Big]^T.
    \end{aligned}
\end{equation}

Under the independent assumption, probabilities $P(\gamma_j|r_j\in M,\Theta)$ and $P(\gamma_j|r_j\in U,\Theta)$ can be computed as $\prod_{i = 1}^{m}P(\gamma_j^{(i)}|r_j\in M,\Theta)$ and $\prod_{i = 1}^{m}P(\gamma_j^{(i)}|r_j\in U,\Theta)$, where $m$ is the number of similarity functions.
However, for different similarity functions, the distributions of $\gamma_j^{(i)}$ are different from each other.
To simplify the log-likelihood, we employ the exponential family to model the distribution of similarity $\gamma_j^{(i)}$.

In the above probabilistic generative model, $\mathbf{l}^j$ is a random vector.
Thus, the parameter vector $\Theta$ cannot be directly learned by their MLEs.
IUAD adopts the EM algorithm to learn the MLEs of parameters~\cite{gao2015cnl}.
The MLEs of parameters for distributions are summarized in Table~\ref{table:mles}. 

\begin{table*}[htbp]
\centering
   \caption{\textbf{MLEs of parameters for matched and unmatched groups, where $N$ is the number of samples, $\left \langle k \right \rangle$ represents the $k$-th iteration, and $\cdot_{1,i}, \cdot_{2,i}$ denote the MLEs of matched and unmatched groups for the $i$-th similarity function, respectively.}}
  \begin{tabular}{p{2cm}|p{5.5cm}|p{5.5cm}}
  \toprule
  \textbf{Distribution} & \textbf{MLEs for matched group} & \textbf{MLEs for unmatched group}\\
  \midrule
  \textbf{Multinomial} & $p_{1,i}^{h,	\left \langle k	\right \rangle} = \frac{ \sum_{j=1}^{N}l_j^{	\left \langle k	\right \rangle}I_{\gamma^{(i)}_j=h}}{\sum_{j=1}^{N}l_j^{\left \langle k \right \rangle}}$ 
  & $p_{2,i}^{h,\left \langle k \right \rangle} = \frac{\sum_{j=1}^{N}(1-l_j^{\left \langle k \right \rangle})I_{\gamma^{(i)}_j=h}}{N - \sum_{j=1}^{N}l_j^{\left \langle k \right \rangle}}$\\
  \cmidrule(lr){1-3}
  \multirow{2}{*}{\textbf{Gaussian}}
  & $\mu_{1,i}^{\left \langle k \right \rangle} = \frac{\sum_{j=1}^{N}l_j^{\left \langle k \right \rangle} \gamma^{(i)}_j}{\sum_{j=N}^{N}l_j^{\left \langle k \right \rangle}}$
  & $\mu_{2,i}^{\left \langle k \right \rangle} = \frac{\sum_{j=1}^{N}(1-l_j^{\left \langle k \right \rangle})\gamma^{(i)}_j}{N - \sum_{j=1}^{N}l_j^{\left \langle k \right \rangle}}$ \\
  & $(\sigma_{1,i}^{\left \langle k \right \rangle})^2 = \frac{ \sum_{j=1}^{N}l_j^{\left \langle k \right \rangle}(\gamma^{(i)}_j - \mu_{1,i}^{\left \langle k \right \rangle})^2}{ \sum_{j=1}^{N}l_j^{\left \langle k \right \rangle}}$
  & $(\sigma_{2,i}^{\left \langle k \right \rangle})^2 = \frac{ \sum_{j=1}^{N}(1-l_j^{\left \langle k \right \rangle})(\gamma^{(i)}_j - \mu_{2,i}^{\left \langle k \right \rangle})^2}{N - \sum_{j=1}^{N}l_j^{\left \langle k \right \rangle}}$ \\
  \cmidrule(lr){1-3}
  \textbf{Exponential} & $\lambda_{1,i}^{\left \langle k \right \rangle} = \frac{ \sum_{j=1}^{N}l_j^{\left \langle k \right \rangle}}{ \sum_{j=1}^{N}l_j^{\left \langle k \right \rangle}\gamma^{(i)}_j}$
  & $\lambda_{2,i}^{\left \langle k \right \rangle} = \frac{N - \sum_{j=1}^{N}l_j^{\left \langle k \right \rangle}}{ \sum_{j=N}^{N}(1-l_j^{\left \langle k \right \rangle})\gamma^{(i)}_j} $\\
  \bottomrule
  \end{tabular}
  \label{table:mles}
  \vspace{-0.5cm}
\end{table*}

\subsection{Decision Making}
\label{subsec:outcome_assignment}
Once we learn the parameters of the model, the computed probabilities $P(r_j\in M|\gamma_j,\hat{\Theta})$ and $P(r_j\in U| \gamma_j,\hat{\Theta})$ can be used to judge whether two vertices in SCN are a unique author or not.
To simplify the judgment process, IUAD defines the following matching score for vertex pair $r_j$:
\begin{equation}
    \begin{aligned}
    sc_j = \log\big(\frac{P(r_j \in M|\gamma_j,\hat{\Theta})}{P(r_j \in U|\gamma_j,\hat{\Theta})}\big).
    \end{aligned}
\label{equ:score_function}
\end{equation}
When $sc_j$ is much larger than 0, two vertices in $r_j$ are more likely to be merged as a single vertex.
As shown in Algorithm~\ref{alg:IUAD}, given decision threshold $\delta$, if $sc_j \geq \delta$, two vertices in $r_j$ are merged.

\subsection{Incremental Manner}
\label{subsec:incremental_manner}
For a newly published paper $p^a$, which is published by author $a$, then we need to judge whether author $a$ is identical to an existing vertex in GCN.
The problem is called the single paper disambiguation problem.
Since all vertices named $a$ are different authors after constructing GCN, our proposed IUAD method can easily extend to incrementally solve the single paper disambiguation problem.

For new paper $p^a$, $a$ is viewed as an isolated vertex, denoted as $v^a$, in the GCN.
Let $V^a=\{v_1^a, v_2^a, \cdots,v_n^a\}$ be the set of vertices named $a$ in GCN.
The single paper disambiguation problem aims at judging whether $v^a$ is identical to $v_k^a\in V^a$ for a fixed  $k$ $(1 \leq k \leq n )$.

To make the decision, we first calculate similarity vector $\gamma_i^a$ for vertex pair $r_i^a = (v^a, v_i^a)$.
Based on learned parameter vector $\hat{\Theta}$, we can compute probabilities $P(r_i^a\in M|\gamma_i^a, \hat{\Theta})$ and $P(r_i^a\in U|\gamma_i^a, \hat{\Theta})$.
We then compute matching score $sc_i^a$ for vertex pair $r_i^a$ by using Equation~(\ref{equ:score_function}).
Since $v^a$ is identical at most one vertex $v_k^a$ in $V^a$, we make the decision that $v^a$ is identical to $v_k^a$ if the following two conditions satisfy: (1) for $i\neq k$ and $v_i^a\in V^a$, $sc_k^a\geq sc_i^a$ ; (2)  $sc_k^a\geq \delta$.
That is, $v^a$ is identical to $v_k^a$ if the largest matching score $sc_k^a$ is larger than our pre-defined threshold $\delta$.
If the conditions do not satisfy, $v^a$ will be an isolated vertex in the constructed GCN.
For newly published papers, in contrast to the existing embedding-based approaches that need to re-train the whole model, our proposed IUAD method can incrementally address the single paper disambiguation problem one by one efficiently.

%% file: discussion.tex
\subsection{Discussion}
\label{sec:discuss}

\subsubsection{Efficiency of IUAD} 

IUAD is an efficient algorithm mainly due to the following facts:

\begin{itemize}
    \item In SCN construction stage, it only mines the frequent 2-itemsets which can be efficiently found, instead of using high complexity operations such as extracting paths or cycles from the network. Furthermore, IUAD captures the higher-order collaborative relations via inferring the stable triangles from the found $\eta$-SCRs. In this stage, IUAD therefore performs efficiently.

    \item In GCN construction stage, we only merge the vertices in the SCN, which share the same name. Note that a small number of names are shared by many vertices in SCN. To further reduce the complexity of GCN construction, we randomly sample a small number of vertex pairs (i.e., 10\%), which share the same name, to train the generative probabilistic model. The sampling strategy speeds up the computation significantly.
\end{itemize}
In existing top-down approaches, they involve many repeated calculations. For example, given a paper of five authors, it will be considered five times to generate different ego-networks to disambiguate each author. However, our proposed IUAD method aims at recovering the global collaborative network in the bottom-up manner, and avoids the repeated calculations. Comparing to the existing top-down approaches, our IUAD is therefore more efficient than them.

\subsubsection{Sampling Strategy} 
When we learn parameter vector $\Theta$ of the generative probabilistic model, we encounter the imbalance problem, i.e., there are only a small number of matched vertex pairs comparing with the unmatched vertex pairs. To reduce the negative impact of imbalance problem, we partition a vertex in SCN, which has many published papers, into two vertices at random. As such, the matched and unmatched vertex pairs are more balanced for training when the number of matched vertex pairs increases.

%% file: experiments.tex
\section{EXPERIMENTS}
\label{sec:experiments}
We conduct comprehensive experiments to evaluate the performance and efficiency of IUAD via comparing it with 8 competitors on the real DBLP data. Through empirical study, we aim at addressing the following research questions:
\begin{itemize}
    \item \textbf{RQ1} How does IUAD perform comparing with baselines?
    
    \item \textbf{RQ2} What is the contribution of each stage of IUAD?
    
    \item \textbf{RQ3} Does IUAD perform well on the large-scale dataset?
    
    \item \textbf{RQ4} Whether is the incremental manner practicable in performance and efficiency?
    
    \item \textbf{RQ5} Whether are similarity functions reasonable?
\end{itemize}

\subsection{Experimental Settings}
\label{subsec:experimental_settings}

\subsubsection{\textbf{Dataset}}

We use a large-scale DBLP dataset\footnote{https://dblp.uni-trier.de/xml/} to conduct our experiments. In this dataset, there are 72,522 different author names, 641,377 published papers associated with co-author lists, paper titles, venues, and published years.
Totally, there are 2,393,969 author-paper pairs.

To evaluate the performance of IUAD and competitors, we build a testing dataset via 
intersecting two datasets DBLP and DAminer~\cite{zhang2018name}\footnote{https://github.com/neozhangthe1/disambiguation/}.
We do not evaluate the performance of IUAD and competitors on dataset DAminer since 
it only contains papers published by randomly selecting 600 author names, which only support to partially preserve the collaborative relations for the selected author names.
Furthermore, IUAD method cannot fully capture the collaborative relations between authors on dataset DAminer.
As a result, we obtain a testing dataset, which contains 336 real authors and 50 different author names who co-exist in both datasets DBLP and DAminer.
Table~\ref{table:testinhg_dataset} lists the descriptive statistics of 50 different author names in our testing dataset, where \#Authors\_TD and \#Papers\_TD are the number of authors and the number of papers in the testing dataset, respectively, and \#Papers\_DBLP denotes the number of papers in the whole DBLP dataset.

\begin{table*}[tbp]
\centering
  \caption{\textbf{The descriptive statistics for our testing dataset.}}
  \begin{tabular}{p{1.7cm}<{\centering}p{1.45cm}<{\centering}p{1.45cm}<{\centering}p{1.9cm}<{\centering}|p{1.9cm}<{\centering}p{1.45cm}<{\centering}p{1.45cm}<{\centering}p{1.8cm}<{\centering}}
  \toprule
  \textbf{Name} & \textbf{\#Authors\_TD} & \textbf{\#Papers\_TD} & \textbf{\#Papers\_DBLP} & \textbf{Name} & \textbf{\#Authors\_TD} & \textbf{\#Papers\_TD} & \textbf{\#Papers\_DBLP} \\
  \midrule
  Jia Xu & 12 & 79  & 223 & Song Chen & 5 & 49 & 121 \\
  Lixin Tang& 4 & 79 & 112 & Bo Ai & 4 & 98 & 179 \\
  Ping Fu& 5 & 16  & 32 & Wensheng Yang & 4 & 14 & 14 \\
  Qi Hu & 5 & 9 & 33 & Hongbin Liang & 3 & 5 & 19 \\
  Geng Yang & 4 & 55 & 92 & Yang Shen & 14 & 30 & 65 \\
  Xu Xu & 12 & 60 & 110 & Jing Luo & 8 & 35 & 61 \\
  Jianhua Lu & 7 & 138 & 222 & Jian Du & 5 & 15 & 41 \\
  Lin Huang & 12 & 75 & 105 & Lu Han & 11 & 29 & 57 \\
  Yong Tian & 10 & 13 & 25 & Rong Yu & 4 & 53 & 101 \\
  Jian Feng & 9 & 33 & 85 & Bo Hong & 5 & 18 & 78 \\
  Wei Quan & 4 & 32 & 87 & Tao Deng & 7 & 10 & 33 \\
  Hongbin Li & 8 & 65 & 258 & Yun Zhou & 16 & 41 & 137 \\
  Hua Bai & 4 & 9 & 16 & Yanqing Wang & 6 & 17 & 47 \\
  Ping Sun & 9 & 14 & 37 & Jianqiang Yi & 3 & 84 & 124 \\
  Dandan Zhang & 8 & 26 & 39 & Weiwei Li & 10 & 22 & 66 \\
  Xi Huang & 7 & 17 & 26 & Xue Qin & 3 & 4 & 6 \\
  Jie Jiang & 8 & 39 & 109 & Fei Sun & 8 & 17 & 58 \\
  Lei Song & 17 & 44 & 118 & Junling Wang & 2 & 5 & 15\\
  Shuai Yuan & 8 & 18 & 96 & Tian Chen & 6 & 10 & 20 \\
  Min Zheng & 6 & 15 & 42 & Chuanyan Liu & 6 & 9 & 18 \\
  Minghui Li & 6 & 14 & 32 & Yin Shi & 2 & 11 & 17 \\
  Zhifeng Liu & 5 & 12 & 20 & Rong Lu & 3 & 4 & 11 \\
  Hongtao Liu & 5 & 13 & 21 & Hong Fan & 6 & 25 & 50 \\
  Yin Wu & 9 & 15 & 38 & Shuang Song & 3 & 9 & 52 \\
  Dan Sun & 4 & 9 & 18 & Lili Ma & 4 & 16 & 40 \\
  \midrule
  Total  & 336 & 1529 & 3426 &  & & & \\
  \bottomrule
  \end{tabular}
  \label{table:testinhg_dataset}
\end{table*}

\subsubsection{\textbf{Evaluation Metrics.}}

To avoid the disturbance of different numbers of papers published by different authors, micro-accuracy (MicroA), micro-precision (MicroP), micro-recall (MicroR), and micro-F1 (MicroF) are utilized to evaluate the performance of IUAD and its competitors. These measurements can be computed as followings:
\begin{align*}
MicroA &= \frac{TP + TN}{ TP+FP+FN+TN}, \\ 
MicroP &= \frac{TP}{ TP + FP}, \\ 
MicroR &=  \frac{TP}{ TP + FN},\\ 
MicroF &=  \frac{2 * MicroP * MicroR}{MicroP + MicroR},
\end{align*}
where \textit{TP} and \textit{FP} are the numbers of paper pairs that they are correctly and incorrectly predicted to be from the same author, respectively; \textit{FN} and \textit{TN} are the number of paper pairs that the they are incorrectly and correctly predicted to be from different authors, respectively. 
To reduce the impact of the imbalance in published papers for different names,
all the values of \textit{TP}, \textit{FP}, \textit{FN}, and \textit{TN} count the total number of corresponding paper pairs of all names.

For evaluating the efficiency of methods, the average time cost per name is also computed.

\subsubsection{\textbf{Baselines}}

We compare IUAD with unsupervised and supervised baselines. For comparing baselines, their experimental settings are consistent with their original papers. For our IUAD, we only sample 10\% vertex pairs to training the generative probabilistic model in the stage of GCN construction.

\begin{enumerate}
\item[(i)] \textbf{Unsupervised Baselines}
Similar to IUAD, there is a set of unsupervised methods~\cite{zhang2017name, xu2018network, zhang2018name, fan2011graph} to disambiguate authors, where NetE achieves the state-of-the-art performance~\cite{xu2018network}. 
    \begin{itemize}
        \item \textbf{ANON}~\cite{zhang2017name}: ANON employs the Hierarchical Agglomerative Clustering (HAC) to cluster papers after embedding papers into a low-dimensional space, where all papers in a cluster are published by a unique author.
        
        \item \textbf{NetE}~\cite{xu2018network}: NetE embeds papers into a low-dimensional space via mining multiple relationships. Furthermore, NetE employs HDBSCAN and AP (i.e., Affinity Propagation) methods to cluster papers.
        
        \item  \textbf{Aminer}~\cite{zhang2018name}: Aminer leverages both global and local information to embed papers into a low-dimensional space. To improve disambiguation accuracy, Aminer also leverages human annotations to learn paper embeddings. 
        Furthermore, Aminer disambiguates authors via employing the HAC algorithm to group papers.

        \item \textbf{GHOST}~\cite{fan2011graph}: GHOST is a graph-based method, which devises a path-based similarity metric to evaluate the similarity between papers, and further groups papers into clusters with the AP algorithm. 
    \end{itemize}

\item[(ii)] \textbf{Supervised Baselines}
To verify the effectiveness of our method, we compare IUAD with the supervised methods. We employ AdaBoost, GBDT, RF, and XGBoost to learn classifiers, which are to determine whether two papers are published by a unique author or not. For extracting features, we follow the work of Treeratpituk et al.~\cite{treeratpituk2009disambiguating}.
\end{enumerate}

\subsection{Performance Comparison (RQ1)}

\begin{table}[tbp]
    \centering
    \caption{\textbf{Performance compared with baselines.}}\label{tab:exp_compare_baselines}
 \begin{tabular}{p{1.4cm}<{\centering}|p{1.2cm}<{\centering}|p{0.95cm}<{\centering}|p{0.95cm}<{\centering}|p{0.95cm}<{\centering}|p{0.95cm}<{\centering}}
        \toprule
        \multicolumn{2}{c|}{ \textbf{Algorithm} } & \textbf{MicroA} & \textbf{MicroP} & \textbf{MicroR} & \textbf{MicroF}\\
        \midrule
        \midrule
         \multirow{4}{*}{\textbf{Supervised}}
        & \textbf{AdaBoost}  & 0.6812 &  0.6891 &  0.8046 & 0.7424\\
        & \textbf{GBDT}  & 0.6914& 0.7422 & 0.7041 & 0.7226\\
        & \textbf{RF}  & 0.7118 & 0.7215  & 0.8066 & 0.7617 \\
        & \textbf{XGBoost}  & 0.6935 & 0.7467 & 0.7009 & 0.7231\\
        \midrule
        \midrule
        \multirow{5}{*}{\tabincell{c}{\textbf{Un-} \\ \textbf{Supervised}}}
        & \textbf{ANON} & 0.6697 &0.8164 & 0.5438 & 0.6528\\
        & \textbf{NetE}  & 0.7318  & 0.8273 & 0.6702& 0.7405\\
        & \textbf{Aminer}  & 0.6182& 0.8235 & 0.4217 & 0.5578\\
        & \textbf{GHOST} & 0.4800 & 0.6814 & 0.1675 & 0.2690 \\
        \midrule
        \midrule
        {\textbf{Our}}& \textbf{IUAD} & \textbf{0.8174} & \textbf{0.8608} & \textbf{0.8113} &  \textbf{0.8353}\\
        \bottomrule
    \end{tabular}
\end{table}

To demonstrate the performance of our proposed IUAD method, we compare IUAD with four unsupervised methods and four supervised methods with four evaluating metrics. The experimental results are illustrated in Table~\ref{tab:exp_compare_baselines}, where we have the following key observations:
\begin{itemize}
\item Comparing to supervised methods, IUAD has achieved better performance.
This is due to the fact that supervised approaches do not consider co-author relationships.
This points to the positive effect of capturing the collaborative relations during SCN and GCN construction stages.

\item IUAD outperforms ANON and NetE significantly.
The suboptimal performance of them is due to the fact that ANON and NetE model the co-author relationships as an ego-network, which captures the low-quality collaborative relations.
This sheds light on the benefit of our bottom-up author disambiguation method, which preserves the high-quality collaborative relations in the beginning.

\item IUAD outperforms both Aminer and GHOST significantly. 
Although both Aminer and GHOST preserve higher-order collaborative relationships, their mined higher-order collaborative relationships will also introduce new uncertainty into the task when author disambiguation has not done yet.
This also points to the positive effect of finding stable collaborative relations from the co-author lists.
\end{itemize}

\subsection{Effect Analysis of Different Stages (RQ2)}
\label{subsec:component}

\begin{table}[t]
    \centering
    \caption{\textbf{Effect of two stages.}}
    \label{tab:incremental}
    \begin{tabular}{p{1cm}<{\centering}|p{1.3cm}<{\centering}|p{1.3cm}<{\centering}|p{1.3cm}<{\centering}}
        \toprule
         & \textbf{SCN} & \textbf{GCN} & \textbf{Improv.}\\
        \midrule
        \textbf{MicroA} & 0.6402 &  0.8174 & +0.1772\\
        \midrule
        \textbf{MicroP} &  0.8662 &  0.8608 & -0.0054\\
        \midrule
        \textbf{MicroR} &  0.4374 &   0.8113 & +0.3739\\
        \midrule
        \textbf{MicroF} & 0.5813  &  0.8353 & +0.2540\\
        \bottomrule
    \end{tabular}
\end{table}

IUAD is a two-stage author disambiguation method. 
To demonstrate the rationality of our two-stage method, we evaluate the performance after SCN and GCN construction stages of IUAD as illustrated in Table~\ref{tab:incremental}, where we have the following key observations:
\begin{itemize}
\item IUAD achieves a high precision after the stage of SCN construction. 
It indicates that our proposed way of capturing stable collaborative relations is rather effective to identify authors, and ensures the high-precision of IUAD.

\item After the stage of GCN construction, the improvements compared to the first stage are 17.7\%, -0.5\%, 37.4\%, and 25.4\% on four evaluating metrics. 
There are two paramount findings that: (1) the biggest improvement appears on micro-recall, where the improvement is up to 37.4\%; (2) although we identify more authors in this stage, the precision of IUAD only decreases 0.5\%. 
The experimental result reveals that: (1) our proposed way of constructing the global collaboration network is also effective to disambiguate authors; (2) our proposed two-stage collaboration network reconstruction makes a very good balance in micro-precision and micro-recall.
\end{itemize}

\subsection{Scalability Analysis (RQ3)}
\label{subsec:scalability}

We report the average time cost per name disambiguation of our IUAD and four unsupervised baselines in Table~\ref{tab:time_compare}. 
We can observe that IUAD is the most efficient method comparing to all unsupervised baselines. 
This is due to the facts that: (1) SCN construction is very efficient since mining all $\eta-$SCRs is very efficient; (2) GCN construction is also efficient since only the vertices with the same name need to judge whether they are a unique author or not; (3) existing top-down approaches will consider a paper multiple times when it has multiple coauthors.  

\begin{table}[tbp]
    \centering
    \caption{\textbf{Average time cost per name disambiguation (seconds).}}
    \label{tab:time_compare}
    \begin{tabular}{p{1.4cm}<{\centering}|p{0.92cm}<{\centering}|p{0.92cm}<{\centering}|p{0.92cm}<{\centering}|p{0.92cm}<{\centering}|p{0.92cm}<{\centering}}
        \toprule
        \textbf{Algorithm} & \textbf{20\%} & \textbf{40\%} & \textbf{60\%} & \textbf{80\%} & \textbf{100\%} \\
        \midrule
        \midrule
        \textbf{ANON} & 4.221 &  9.214 & 17.955 & 35.833 & 58.489 \\
        \textbf{NetE} & 16.113 & 21.597 & 24.396 & 28.798 & 33.093 \\
        \textbf{Aminer} & 2.901 & 3.564 & 4.420 & 5.258 & 6.078 \\
        \textbf{GHOST} & 8.500&21.575 & 44.195&92.165& 183.480 \\
        \midrule
        \midrule
        \textbf{IUAD} & 0.092 & 0.420&1.132 &2.044 & 2.599  \\
        \bottomrule
    \end{tabular}
\end{table}

\begin{figure}[tbp]
  \centering
  \includegraphics[width=0.95\linewidth]{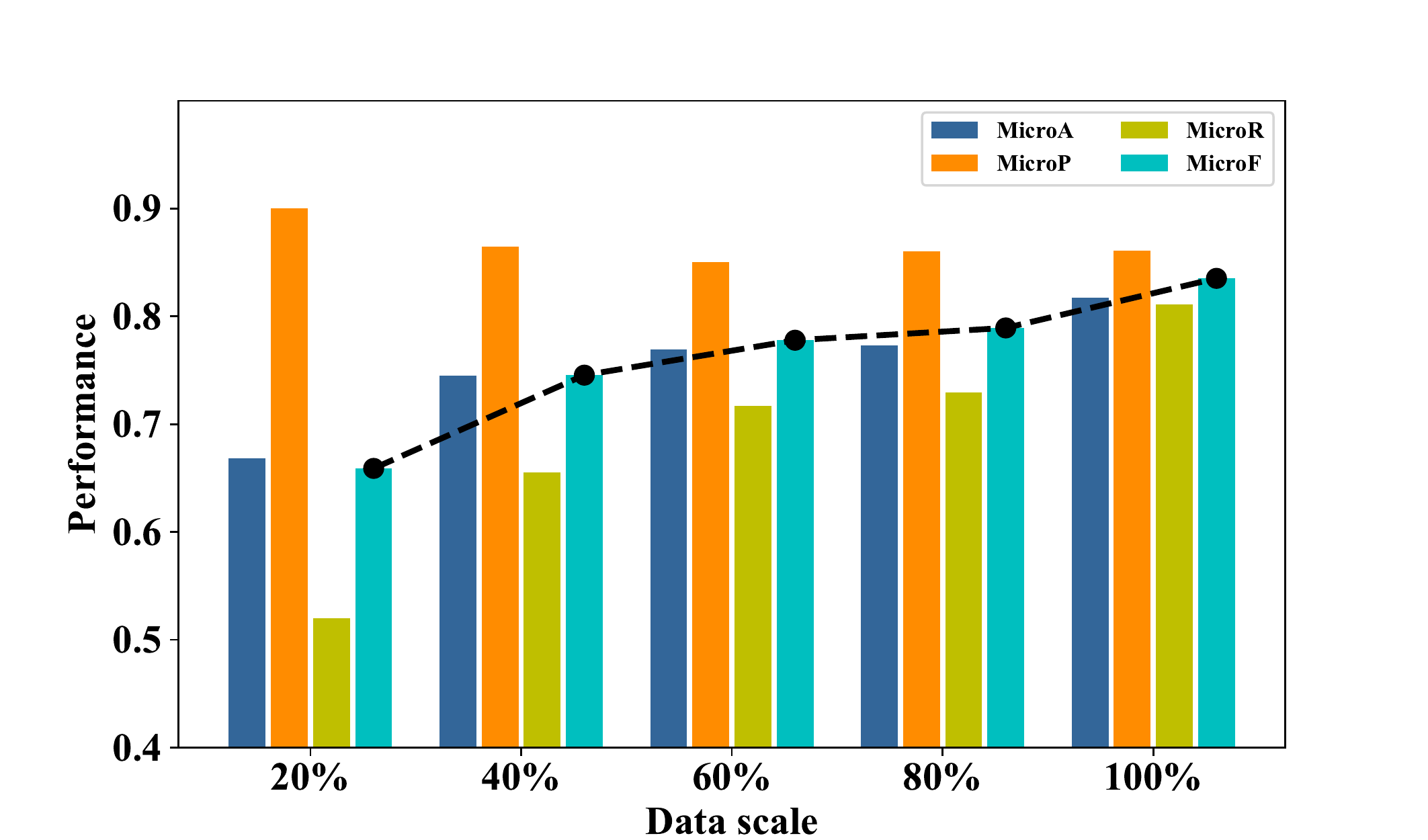}
  \caption{\textbf{Data scale analysis.}}
  \label{fig:scale}
\end{figure}

To further evaluate the scalability of IUAD, we use only 20\%, 40\%, 60\%, 80\% and full data to run IUAD.
The performance is illustrated in Figure~\ref{fig:scale}, where we have the following key observations: (1) IUAD achieves high precision in the SCN construction stage even in a smaller dataset; (2) recall is continually improved from almost 50\% to more than 81\% as data scale increasing.
The observations reveal that: (1) the effect of mining stable collaboration relations during SCN construction is positive; (2) a large-scale dataset is more helpful to construct the GCN.

\subsection{Incremental Manner Analysis (RQ4)}
We also study the performance and efficiency of the incremental author disambiguation task. 
In this task, we separate the testing data into two parts.
The first part is utilized to construct the GCN, and then adopt the second part to study the performance of incremental author disambiguation.
To evaluate the incremental author disambiguation task, the second part of the testing data consists of 100, 200 and 300 papers, which are treated as the recently published ones.
As illustrated in Table~\ref{tab:incremental_manner}, we report the performance and improvement after addressing the incremental author disambiguation problem. In Table~\ref{tab:incremental_manner}, ``MicroA" and ``MicroA+" are the micro-accuracy on the first part of testing data and the entire testing data after incrementally identifying authors, respectively; ``Improv." is the improvement after the incremental author disambiguation comparing to the first stage, other effective metrics are similar, and “Avg. time” is the average elapsed time per paper in the incremental author disambiguation.
We have the following key observations:
\begin{itemize}
\item We almost observe a performance reduction for incremental author disambiguation. 
This is due to the facts that: (1) single paper only provides limited information to identify the authors; (2) the stable collaboration relations in newly published papers are not utilized to disambiguate authors. 
However, incremental author disambiguation does not greatly reduce the performance of our IUAD. 
It indicates that our proposed similarity functions are rational and effective to disambiguate authors.

\item To address the incremental author disambiguation problem, IUAD is very efficient since the average time cost per paper is less than 50 milliseconds. 
This result thanks to: (1) IUAD incrementally judges who publish a newly published paper via only computing the posterior probability, rather than re-training the entire model; (2) once we compute the six similarity functions, we can efficiently calculate the posterior probability.  
\end{itemize}

\begin{table}[tbp]
    \centering
    \caption{\textbf{Performance and efficiency of incremental author disambiguation.}}\label{tab:incremental_manner}
    \begin{tabular}{p{0.95cm}<{\centering}|p{1.1cm}<{\centering}|p{1.25cm}<{\centering}|p{1.25cm}<{\centering}|p{1.25cm}<{\centering}}
        \toprule
        \multicolumn{2}{c|}{ \textbf{Metric} } & \textbf{100} & \textbf{200} & \textbf{300}\\
        \midrule
        \midrule
         \multirow{3}{*}{\textbf{MicroA}}
        & \textbf{MicroA}  & 0.8154 &  0.8104 &  0.8166\\
        & \textbf{MicroA+}  & 0.8062& 0.8079 & 0.8085 \\
        & \textbf{Improv.}  & -0.0092 & -0.0025  & -0.0081  \\
        \midrule
        \midrule
         \multirow{3}{*}{\textbf{MicroP}}
        & \textbf{MicroP}  & 0.8685 &  0.8546 &  0.8544\\
        & \textbf{MicroP+}  & 0.8649& 0.8588 & 0.8606 \\
        & \textbf{Improv.}  & -0.0036 & +0.0042  & +0.0062 \\
        \midrule
        \midrule
        \multirow{3}{*}{\textbf{MicroR}}
        & \textbf{MicroR}  & 0.7974 &  0.8008 &  0.8160\\
        & \textbf{MicroR+}  & 0.7829& 0.7941 & 0.7931 \\
        & \textbf{Improv.}  & -0.0145 & -0.0067  & -0.0229 \\
        \midrule
        \midrule
        \multirow{3}{*}{\textbf{MicroF}}
        & \textbf{MicroF}  & 0.8315 &  0.8268 &  0.8348\\
        & \textbf{MicroF+}  & 0.8218& 0.8252 & 0.8255 \\
        & \textbf{Improv.}  & -0.0097 & -0.0016  & -0.0093 \\
        \midrule
        \midrule
        \multicolumn{2}{c|}{ \textbf{Avg. time (ms)}} & 47.76 & 45.22 & 45.40 \\
        \bottomrule
    \end{tabular}
\end{table}

\subsection{Rationality of Similarity Functions (RQ5)}

\begin{figure}[tbp]
      \centering
      \subfigure[\textbf{Similarity of Representative Community}]{\includegraphics[width=0.49\linewidth]{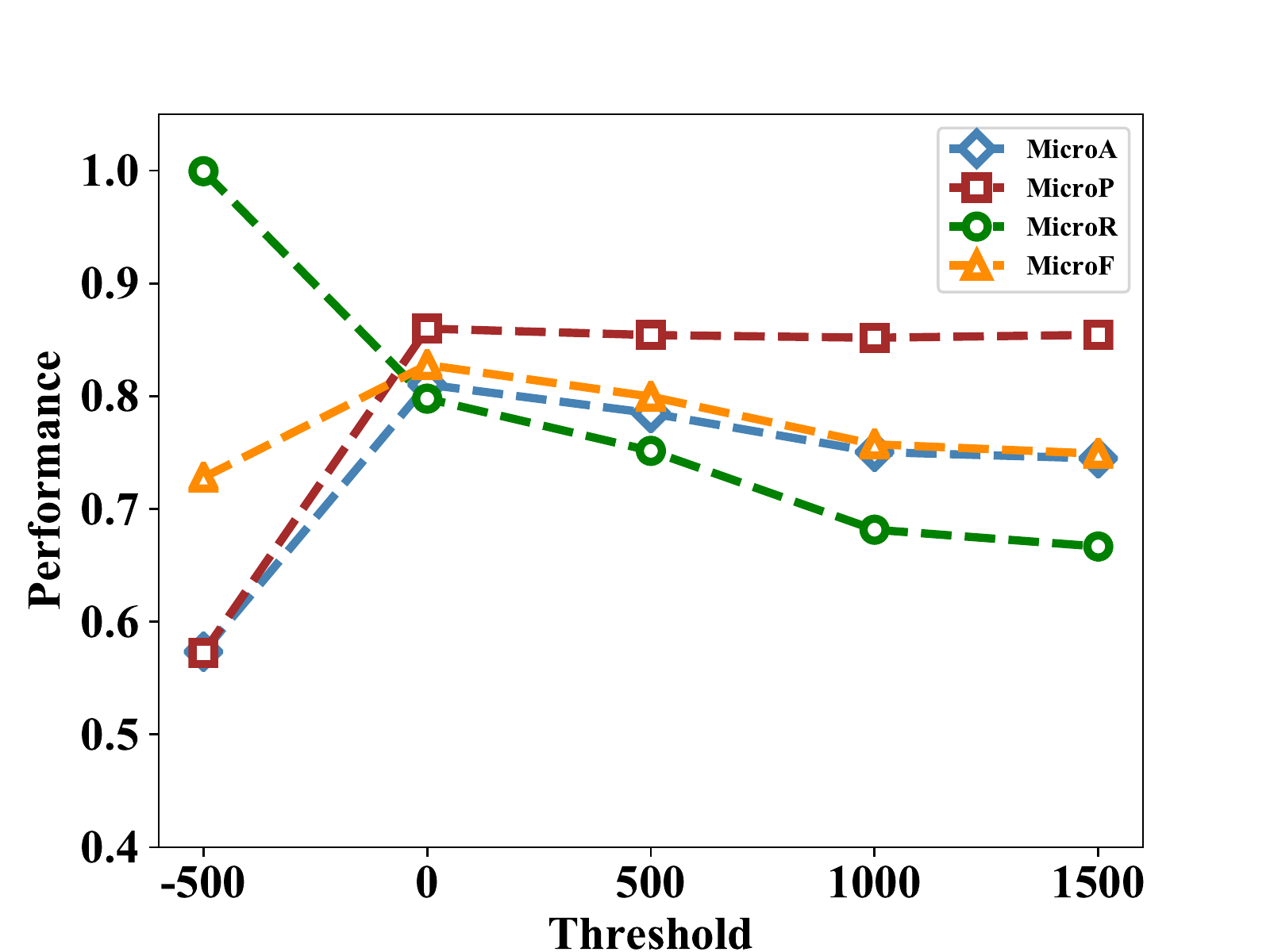}}
      \subfigure[\textbf{Similarity of Research Community}]{\includegraphics[width=0.49\linewidth]{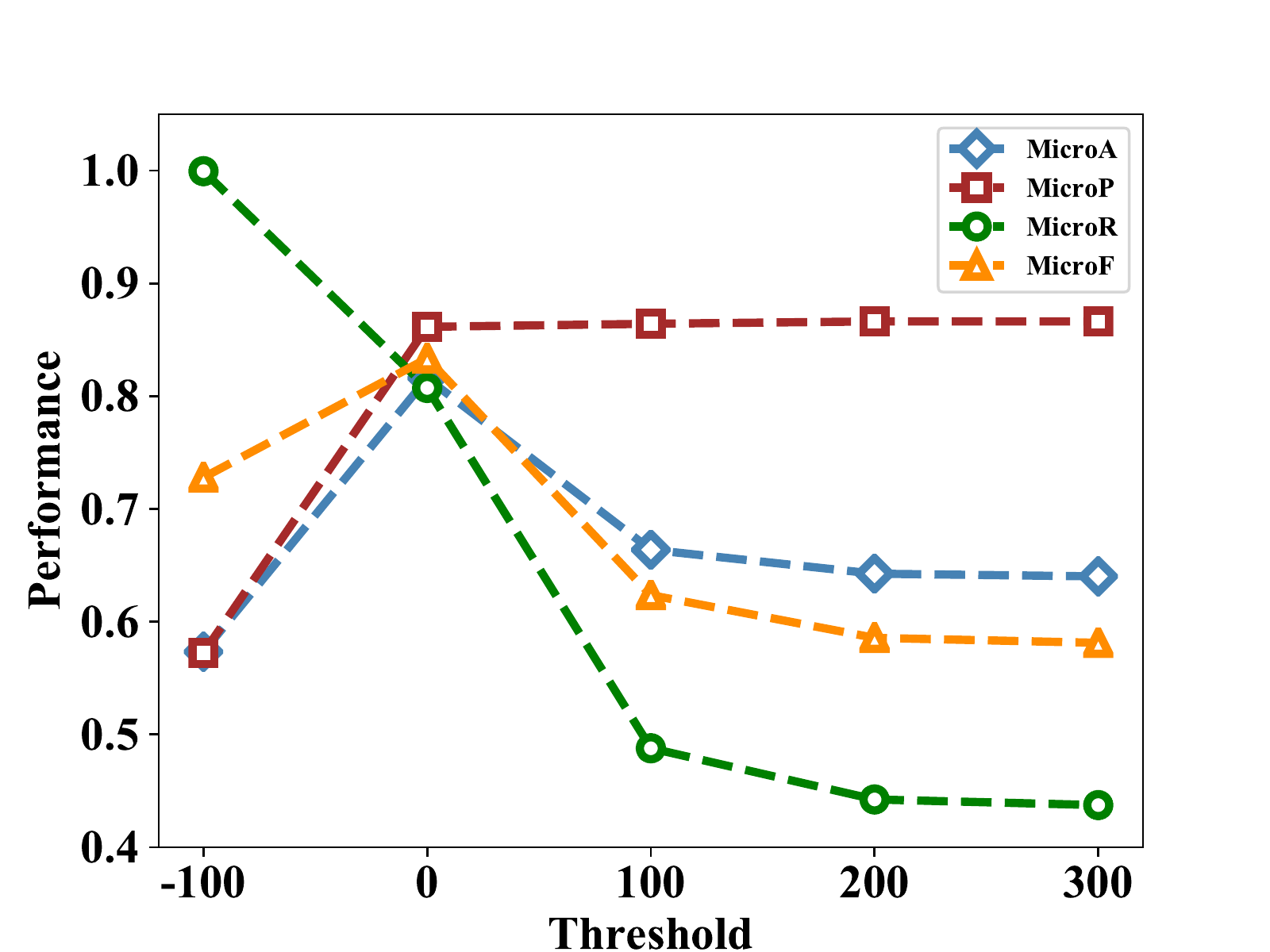}}
      \subfigure[\textbf{Time consistency in research interest}]{\includegraphics[width=0.49\linewidth]{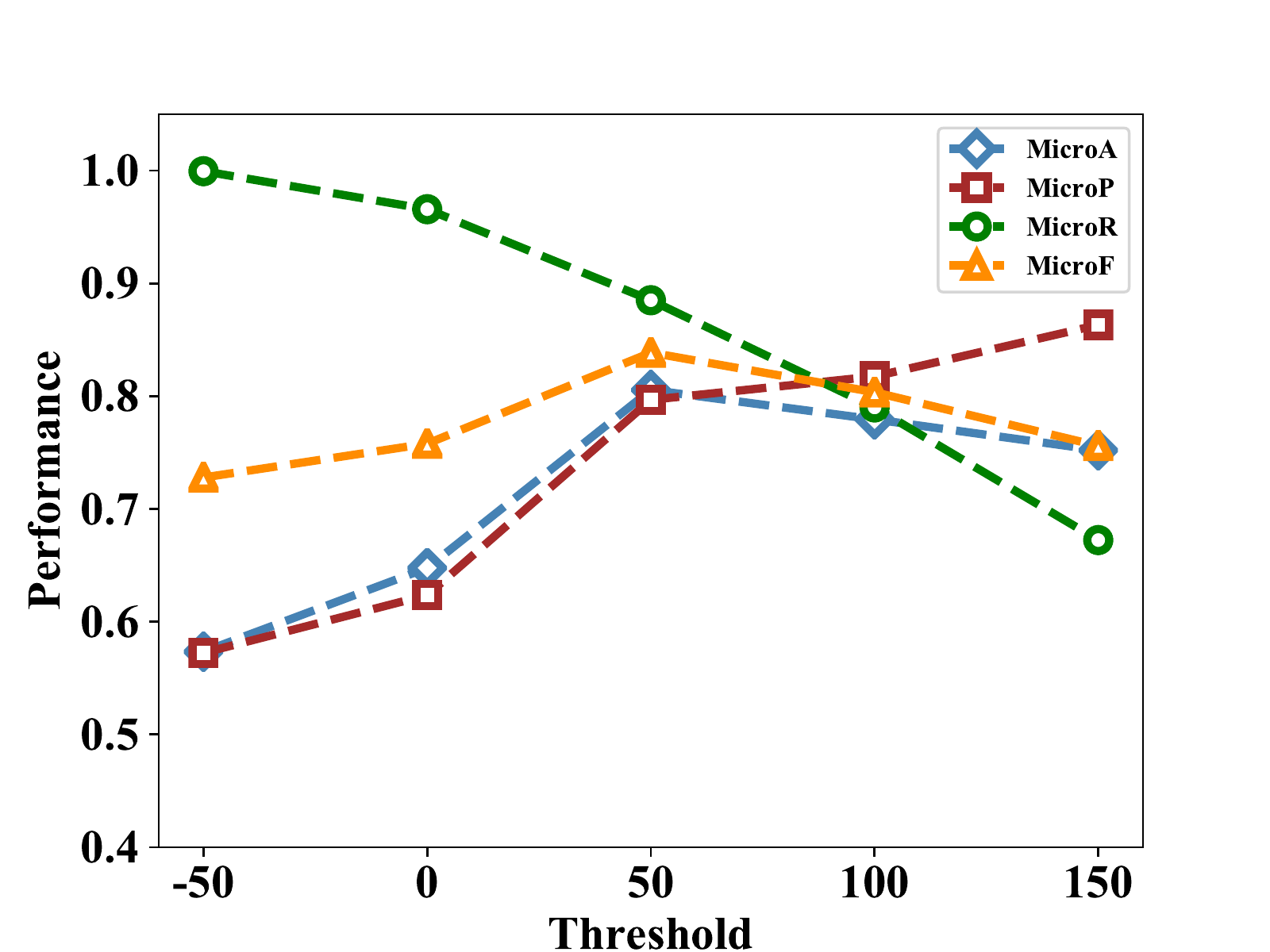}}
       \subfigure[\textbf{Co-author Cliques Coincidence Ratio}]{\includegraphics[width=0.49\linewidth]{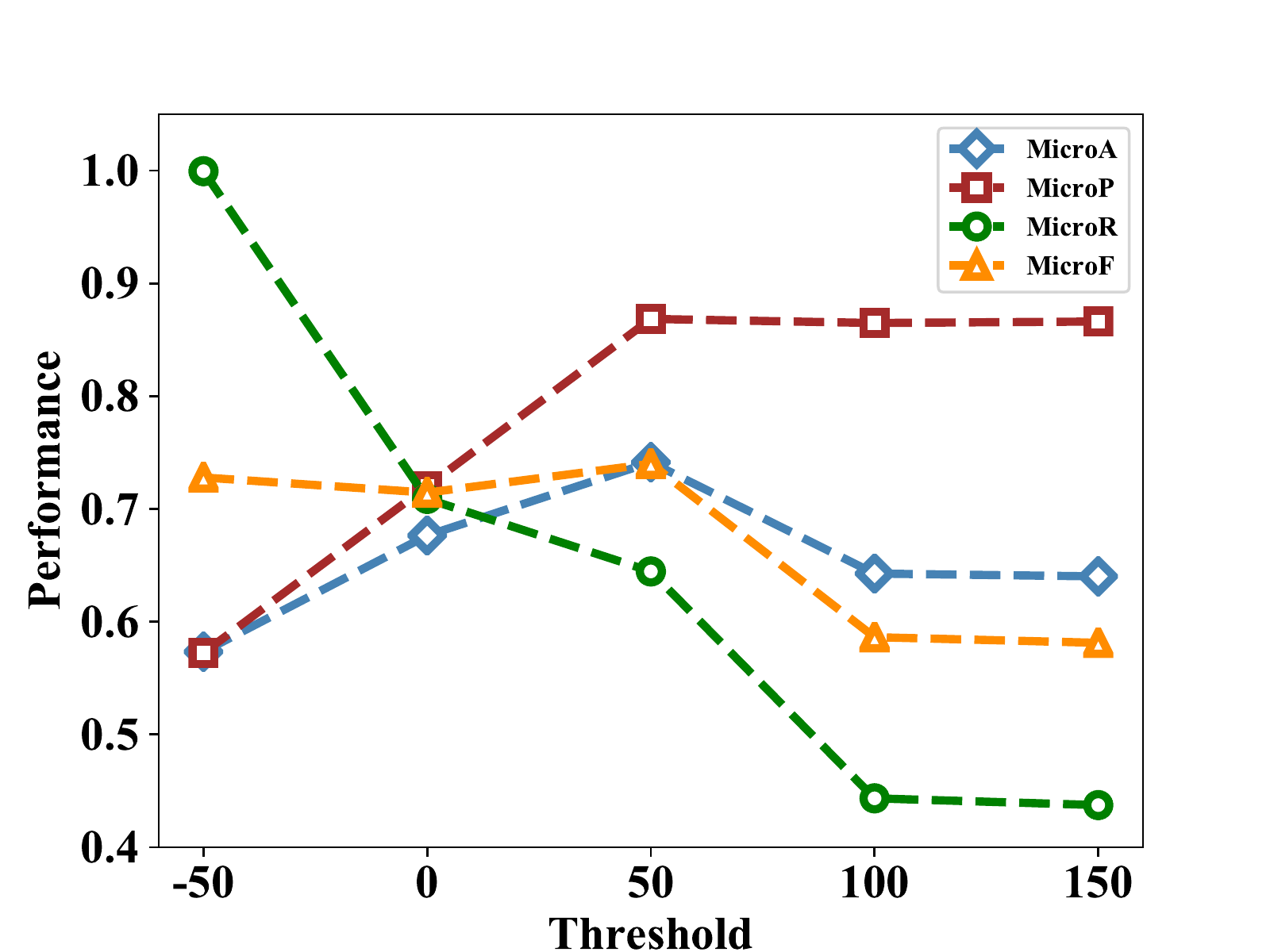}}
      \subfigure[\textbf{Normalized WL Sub-graph Kernel}]{\includegraphics[width=0.49\linewidth]{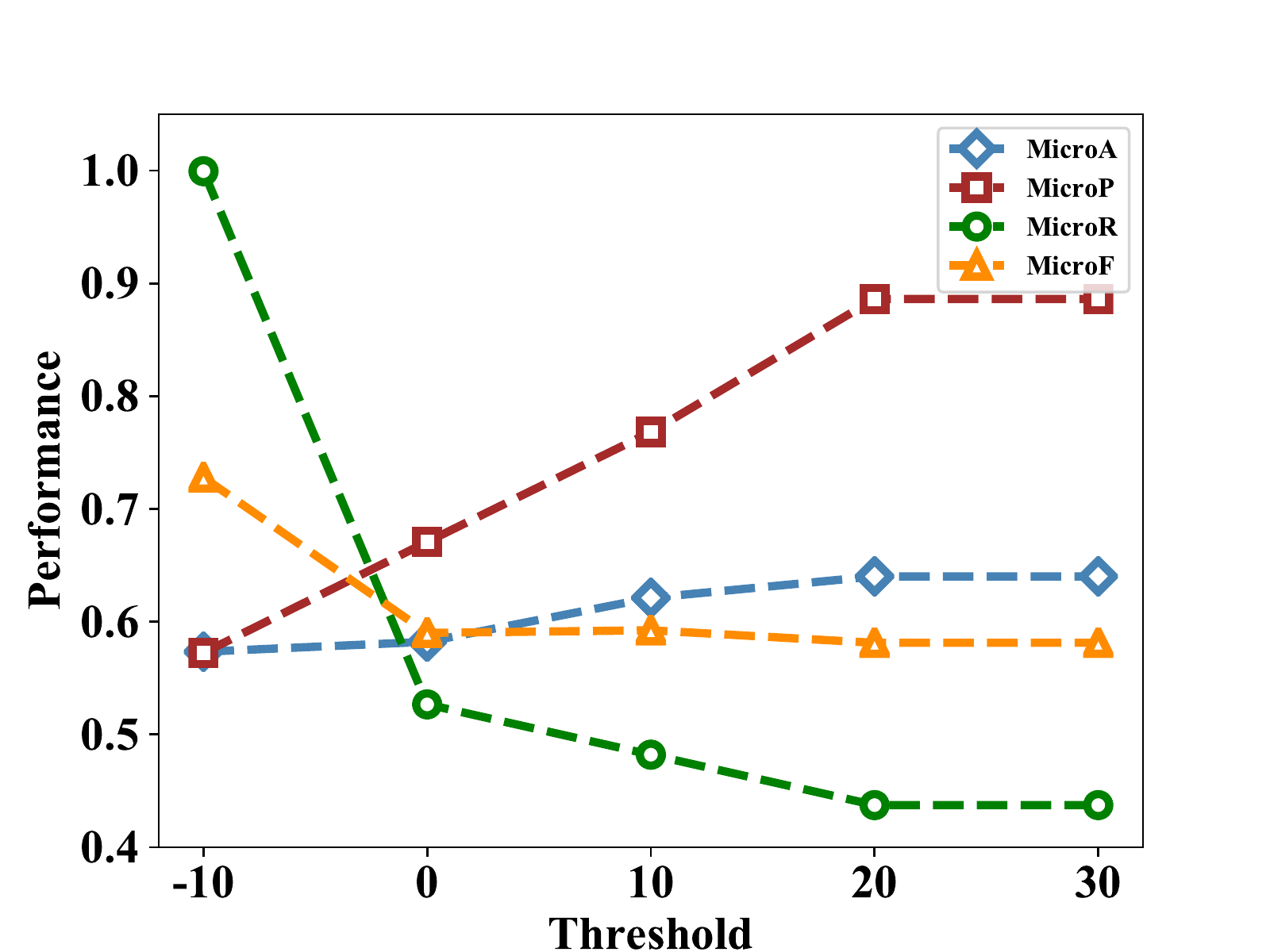}}
      \subfigure[\textbf{Similarity of research interests}]{\includegraphics[width=0.49\linewidth]{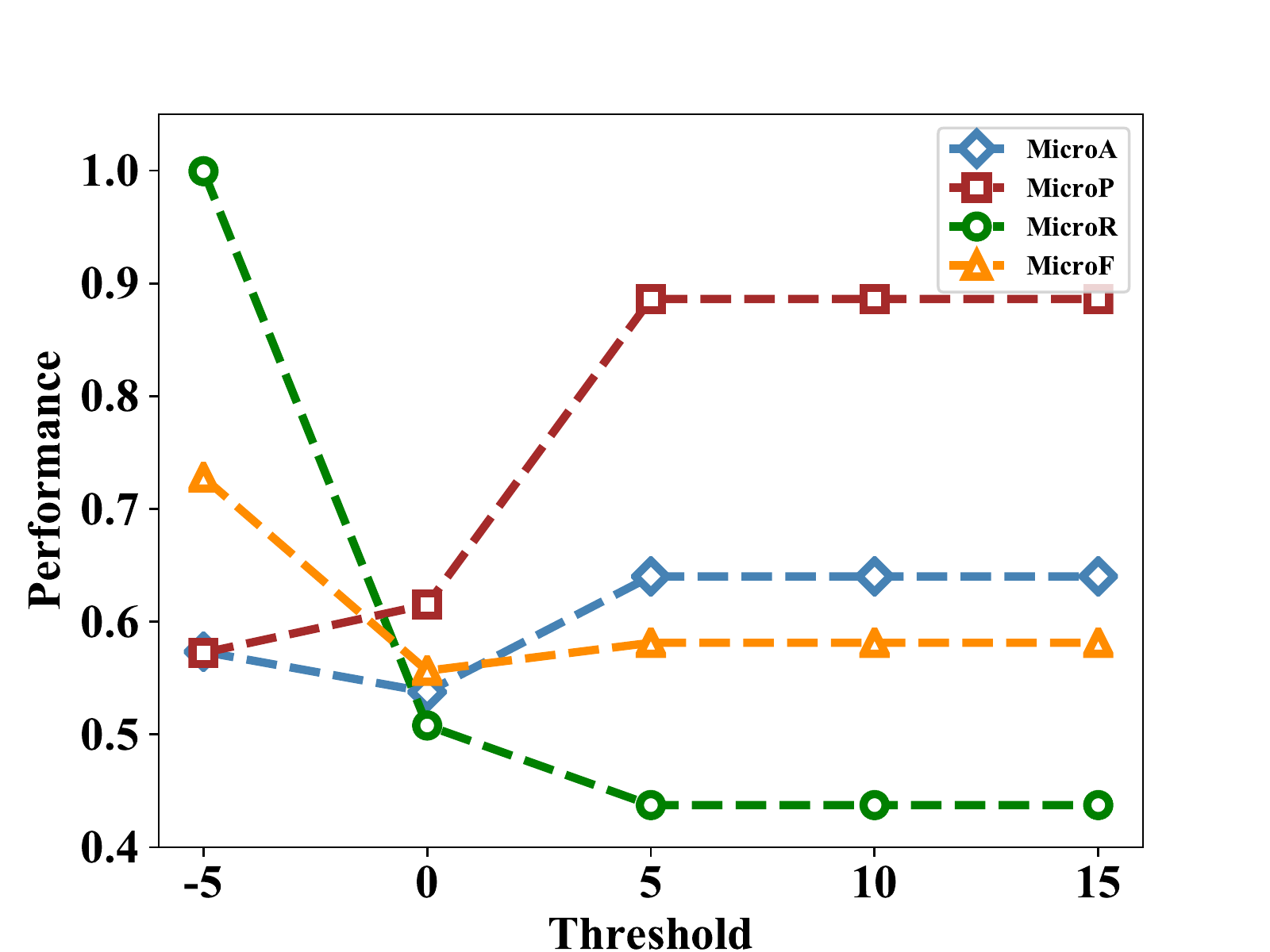}}
      \caption{\textbf{Rationality of similarity functions}}
      \label{fig:feature_analysis}
    \end{figure}

There are six similarity functions designed for recovering GCN in subsection~\ref{subsec:similiarity_functions}. 
To demonstrate the rationality of our proposed six similarity functions, we only employ a single similarity to construct GCN, and illustrate the performance in Figure~\ref{fig:feature_analysis}, where we have the following key observations:
\begin{itemize}
\item We observe that all similarity functions have influences on the performance of IUAD positively. The observation indicates that all similarity functions are reasonable to identify authors.

\item Note that a similarity function is more influential for recovering GCN if its threshold has larger degree of dispersion. Thus, the similarities of representative community (in Figure~\ref{fig:feature_analysis}(a)) and research community (in Figure~\ref{fig:feature_analysis}(b)) are the two most important similarity functions for constructing GCN. This is due to the fact that the stable collaboration relationships have been explored in the SCN construction stage, then the similarities of topological structures (in Figure~\ref{fig:feature_analysis}(d-e)) only capture few helpful information. The time consistency in research interests (in Figure~\ref{fig:feature_analysis}(c)) is more influential than research interests (in Figure~\ref{fig:feature_analysis}(f)), which are only represented by the paper titles.
\end{itemize}

%% file: conclusion.tex
\section{CONCLUSION AND FUTURE WORK}
\label{sec:conclusion}
In this paper, we address the author disambiguation problem in a bottom-up manner.
We design an incremental and unsupervised algorithm that consists of two stages: stable collaboration network construction and global collaboration network construction.
We first construct the stable collaboration network of authors by mining frequent collaborative relations, ensuring the high precision of our proposed IUAD method. 
We then build a probabilistic generative model by employing the exponential family to incorporate six reasonable similarity functions.
Lastly, we develop the EM algorithm to infer the parameters and judge whether two vertices belong to a unique author or not. 
For newly published papers, IUAD can incrementally and efficiently judge who publish these papers in the global collaboration network.
Empirical results demonstrate the effectiveness and efficiency of IUAD and its superiority over the state-of-the-art methods.

In this work, we have addressed the author disambiguation problem in an unsupervised manner, while ignoring the possible labeled data. 
To this end, we plan to extend our method to build a semi-supervised approach to further improve the performance. 
In addition, we are interested in generalizing IUAD to disambiguate entities in the other domains, such as record linkage in the database, entity resolution in knowledge engineering, and de-duplication in data cleaning, etc.

%% file: acknowledgement.tex
\section{Acknowledgement}
This work has been supported by the National Key Research and Development Program of China under grant 2016YFB1000905,
and the National Natural Science Foundation of China under Grant No. U1811264,
U1911203,
61877018,
61672234,
61672384,
the Shanghai Agriculture Applied Technology Development Program, China (Grant No.T20170303),
Shanghai Key Laboratory of Pure Mathematics and Mathematical Practice (Grant No. 18dz2271000), and the Fundamental Research Funds for the Central Universities.

%% file: mybib.tex
\bibliographystyle{IEEEtran}
\bibliography{IEEEabrv,mybib}

%% file: main.bbl
\begin{thebibliography}{10}
\providecommand{\url}[1]{#1}
\csname url@samestyle\endcsname
\providecommand{\newblock}{\relax}
\providecommand{\bibinfo}[2]{#2}
\providecommand{\BIBentrySTDinterwordspacing}{\spaceskip=0pt\relax}
\providecommand{\BIBentryALTinterwordstretchfactor}{4}
\providecommand{\BIBentryALTinterwordspacing}{\spaceskip=\fontdimen2\font plus
\BIBentryALTinterwordstretchfactor\fontdimen3\font minus
  \fontdimen4\font\relax}
\providecommand{\BIBforeignlanguage}[2]{{%
\expandafter\ifx\csname l@#1\endcsname\relax
\typeout{** WARNING: IEEEtran.bst: No hyphenation pattern has been}%
\typeout{** loaded for the language `#1'. Using the pattern for}%
\typeout{** the default language instead.}%
\else
\language=\csname l@#1\endcsname
\fi
#2}}
\providecommand{\BIBdecl}{\relax}
\BIBdecl

\bibitem{winkler1999state}
W.~E. Winkler, ``The state of record linkage and current research problems,''
  in \emph{Statistical Research Division, US Census Bureau}.\hskip 1em plus
  0.5em minus 0.4em\relax Citeseer, 1999.

\bibitem{winkler2006overview}
------, ``Overview of record linkage and current research directions,'' in
  \emph{Bureau of the Census}.\hskip 1em plus 0.5em minus 0.4em\relax Citeseer,
  2006.

\bibitem{fellegi1969theory}
I.~P. Fellegi and A.~B. Sunter, ``A theory for record linkage,'' \emph{Journal
  of the American Statistical Association}, vol.~64, no. 328, pp. 1183--1210,
  1969.

\bibitem{bhattacharya2006latent}
I.~Bhattacharya and L.~Getoor, ``A latent dirichlet model for unsupervised
  entity resolution,'' in \emph{Proceedings of the 2006 SIAM International
  Conference on Data Mining}.\hskip 1em plus 0.5em minus 0.4em\relax SIAM,
  2006, pp. 47--58.

\bibitem{agarwal2018dianed}
P.~Agarwal, J.~Str{\"o}tgen, L.~Del~Corro, J.~Hoffart, and G.~Weikum, ``Dianed:
  time-aware named entity disambiguation for diachronic corpora,'' in
  \emph{Proceedings of the 56th Annual Meeting of the Association for
  Computational Linguistics (Volume 2: Short Papers)}, 2018, pp. 686--693.

\bibitem{moon2018multimodal}
S.~Moon, L.~Neves, and V.~Carvalho, ``Multimodal named entity disambiguation
  for noisy social media posts,'' in \emph{Proceedings of the 56th Annual
  Meeting of the Association for Computational Linguistics (Volume 1: Long
  Papers)}, 2018, pp. 2000--2008.

\bibitem{nie2018mention}
F.~Nie, Y.~Cao, J.~Wang, C.-Y. Lin, and R.~Pan, ``Mention and entity
  description co-attention for entity disambiguation,'' in \emph{Thirty-Second
  AAAI Conference on Artificial Intelligence}, 2018.

\bibitem{tejada2001learning}
S.~Tejada, C.~A. Knoblock, and S.~Minton, ``Learning object identification
  rules for information integration,'' \emph{Information Systems}, vol.~26,
  no.~8, pp. 607--633, 2001.

\bibitem{bilenko2003adaptive}
M.~Bilenko and R.~J. Mooney, ``Adaptive duplicate detection using learnable
  string similarity measures,'' in \emph{Proceedings of the ninth ACM SIGKDD
  international conference on Knowledge discovery and data mining}.\hskip 1em
  plus 0.5em minus 0.4em\relax ACM, 2003, pp. 39--48.

\bibitem{xu2017online}
L.~Xu, A.~Pavlo, S.~Sengupta, and G.~R. Ganger, ``Online deduplication for
  databases,'' in \emph{Proceedings of the 2017 ACM International Conference on
  Management of Data}, 2017, pp. 1355--1368.

\bibitem{delgado2014data}
A.~D. Delgado, R.~Mart{\'\i}nez, V.~Fresno, and S.~Montalvo, ``A data driven
  approach for person name disambiguation in web search results,'' in
  \emph{Proceedings of COLING 2014, the 25th International Conference on
  Computational Linguistics: Technical Papers}, 2014, pp. 301--310.

\bibitem{shen2005constraint}
W.~Shen, X.~Li, and A.~Doan, ``Constraint-based entity matching,'' in
  \emph{AAAI}, 2005, pp. 862--867.

\bibitem{li2019adversarial}
C.~Li, S.~Wang, Y.~Wang, P.~Yu, Y.~Liang, Y.~Liu, and Z.~Li, ``Adversarial
  learning for weakly-supervised social network alignment,'' in
  \emph{Proceedings of the AAAI Conference on Artificial Intelligence},
  vol.~33, 2019, pp. 996--1003.

\bibitem{zhang2016pct}
J.~Zhang and P.~S. Yu, ``Pct: partial co-alignment of social networks,'' in
  \emph{Proceedings of the 25th International Conference on World Wide Web},
  2016, pp. 749--759.

\bibitem{liu2016aligning}
L.~Liu, W.~K. Cheung, X.~Li, and L.~Liao, ``Aligning users across social
  networks using network embedding.'' in \emph{Ijcai}, 2016, pp. 1774--1780.

\bibitem{han2004two}
H.~Han, L.~Giles, H.~Zha, C.~Li, and K.~Tsioutsiouliklis, ``Two supervised
  learning approaches for name disambiguation in author citations,'' in
  \emph{Proceedings of the 2004 Joint ACM/IEEE Conference on Digital Libraries,
  2004.}\hskip 1em plus 0.5em minus 0.4em\relax IEEE, 2004, pp. 296--305.

\bibitem{treeratpituk2009disambiguating}
P.~Treeratpituk and C.~L. Giles, ``Disambiguating authors in academic
  publications using random forests,'' in \emph{Proceedings of the 9th
  ACM/IEEE-CS joint conference on Digital libraries}.\hskip 1em plus 0.5em
  minus 0.4em\relax ACM, 2009, pp. 39--48.

\bibitem{hermansson2013entity}
L.~Hermansson, T.~Kerola, F.~Johansson, V.~Jethava, and D.~Dubhashi, ``Entity
  disambiguation in anonymized graphs using graph kernels,'' in
  \emph{Proceedings of the 22nd ACM international conference on Information \&
  Knowledge Management}.\hskip 1em plus 0.5em minus 0.4em\relax ACM, 2013, pp.
  1037--1046.

\bibitem{kim2019hybrid}
K.~Kim, S.~Rohatgi, and C.~L. Giles, ``Hybrid deep pairwise classification for
  author name disambiguation,'' in \emph{Proceedings of the 28th ACM
  International Conference on Information and Knowledge Management}.\hskip 1em
  plus 0.5em minus 0.4em\relax ACM, 2019, pp. 2369--2372.

\bibitem{atarashi2017deep}
K.~Atarashi, S.~Oyama, M.~Kurihara, and K.~Furudo, ``A deep neural network for
  pairwise classification: Enabling feature conjunctions and ensuring
  symmetry,'' in \emph{Pacific-Asia Conference on Knowledge Discovery and Data
  Mining}.\hskip 1em plus 0.5em minus 0.4em\relax Springer, 2017, pp. 83--95.

\bibitem{wang2008name}
F.~Wang, J.~Li, J.~Tang, J.~Zhang, and K.~Wang, ``Name disambiguation using
  atomic clusters,'' in \emph{2008 The Ninth International Conference on
  Web-Age Information Management}.\hskip 1em plus 0.5em minus 0.4em\relax IEEE,
  2008, pp. 357--364.

\bibitem{zhang2017name}
B.~Zhang and M.~Al~Hasan, ``Name disambiguation in anonymized graphs using
  network embedding,'' in \emph{Proceedings of the 2017 ACM on Conference on
  Information and Knowledge Management}.\hskip 1em plus 0.5em minus 0.4em\relax
  ACM, 2017, pp. 1239--1248.

\bibitem{xu2018network}
J.~Xu, S.~Shen, D.~Li, and Y.~Fu, ``A network-embedding based method for author
  disambiguation,'' in \emph{Proceedings of the 27th ACM International
  Conference on Information and Knowledge Management}.\hskip 1em plus 0.5em
  minus 0.4em\relax ACM, 2018, pp. 1735--1738.

\bibitem{tang2012unified}
J.~Tang, A.~C. Fong, B.~Wang, and J.~Zhang, ``A unified probabilistic framework
  for name disambiguation in digital library,'' \emph{IEEE Transactions on
  Knowledge and Data Engineering}, vol.~24, no.~6, pp. 975--987, 2012.

\bibitem{song2007efficient}
Y.~Song, J.~Huang, I.~G. Councill, J.~Li, and C.~L. Giles, ``Efficient
  topic-based unsupervised name disambiguation,'' in \emph{Proceedings of the
  7th ACM/IEEE-CS joint conference on Digital libraries}.\hskip 1em plus 0.5em
  minus 0.4em\relax ACM, 2007, pp. 342--351.

\bibitem{schulz2014exploiting}
C.~Schulz, A.~Mazloumian, A.~M. Petersen, O.~Penner, and D.~Helbing,
  ``Exploiting citation networks for large-scale author name disambiguation,''
  \emph{EPJ Data Science}, vol.~3, no.~1, p.~11, 2014.

\bibitem{fan2011graph}
X.~Fan, J.~Wang, X.~Pu, L.~Zhou, and B.~Lv, ``On graph-based name
  disambiguation,'' \emph{Journal of Data and Information Quality (JDIQ)},
  vol.~2, no.~2, p.~10, 2011.

\bibitem{shin2014author}
D.~Shin, T.~Kim, J.~Choi, and J.~Kim, ``Author name disambiguation using a
  graph model with node splitting and merging based on bibliographic
  information,'' \emph{Scientometrics}, vol. 100, no.~1, pp. 15--50, 2014.

\bibitem{zhang2019author}
W.~Zhang, Z.~Yan, and Y.~Zheng, ``Author name disambiguation using graph node
  embedding method,'' in \emph{2019 IEEE 23rd International Conference on
  Computer Supported Cooperative Work in Design (CSCWD)}.\hskip 1em plus 0.5em
  minus 0.4em\relax IEEE, 2019, pp. 410--415.

\bibitem{peng2019author}
L.~Peng, S.~Shen, D.~Li, J.~Xu, Y.~Fu, and H.~Su, ``Author disambiguation
  through adversarial network representation learning,'' in \emph{2019
  International Joint Conference on Neural Networks (IJCNN)}.\hskip 1em plus
  0.5em minus 0.4em\relax IEEE, 2019, pp. 1--8.

\bibitem{mondal2019graph}
S.~Mondal and J.~Chandra, ``A graph combination with edge pruning-based
  approach for author name disambiguation,'' \emph{Journal of the Association
  for Information Science and Technology}, 2019.

\bibitem{amancio2015topological}
D.~R. Amancio, O.~N. Oliveira~Jr, and L.~d.~F. Costa,
  ``Topological-collaborative approach for disambiguating authors’ names in
  collaborative networks,'' \emph{Scientometrics}, vol. 102, no.~1, pp.
  465--485, 2015.

\bibitem{zhang2018name}
Y.~Zhang, F.~Zhang, P.~Yao, and J.~Tang, ``Name disambiguation in aminer:
  Clustering, maintenance, and human in the loop,'' in \emph{Proceedings of the
  24th ACM SIGKDD International Conference on Knowledge Discovery \& Data
  Mining}.\hskip 1em plus 0.5em minus 0.4em\relax ACM, 2018, pp. 1002--1011.

\bibitem{liu2015fast}
Y.~Liu, W.~Li, Z.~Huang, and Q.~Fang, ``A fast method based on multiple
  clustering for name disambiguation in bibliographic citations,''
  \emph{Journal of the Association for Information Science and Technology},
  vol.~66, no.~3, pp. 634--644, 2015.

\bibitem{erdos1959random}
P.~Erd{\"o}s and A.~R{\'e}nyi, ``On random graphs, i,'' \emph{Publicationes
  Mathematicae (Debrecen)}, vol.~6, pp. 290--297, 1959.

\bibitem{Tsourakakis08}
\BIBentryALTinterwordspacing
C.~E. Tsourakakis, ``Fast counting of triangles in large real networks without
  counting: Algorithms and laws,'' in \emph{ICDM 2008}, 2008, pp. 608--617.
  [Online]. Available: \url{https://doi.org/10.1109/ICDM.2008.72}
\BIBentrySTDinterwordspacing

\bibitem{han2000mining}
J.~Han, J.~Pei, and Y.~Yin, ``Mining frequent patterns without candidate
  generation,'' in \emph{ACM sigmod record}, vol.~29, no.~2.\hskip 1em plus
  0.5em minus 0.4em\relax ACM, 2000, pp. 1--12.

\bibitem{gao2015cnl}
M.~Gao, E.-P. Lim, D.~Lo, F.~Zhu, P.~K. Prasetyo, and A.~Zhou, ``Cnl:
  collective network linkage across heterogeneous social platforms,'' in
  \emph{2015 IEEE International Conference on Data Mining}.\hskip 1em plus
  0.5em minus 0.4em\relax IEEE, 2015, pp. 757--762.

\bibitem{shervashidze2011weisfeiler}
N.~Shervashidze, P.~Schweitzer, E.~J.~v. Leeuwen, K.~Mehlhorn, and K.~M.
  Borgwardt, ``Weisfeiler-lehman graph kernels,'' \emph{Journal of Machine
  Learning Research}, vol.~12, no. Sep, pp. 2539--2561, 2011.

\bibitem{ah2010normalized}
J.~Ah-Pine, ``Normalized kernels as similarity indices,'' in \emph{Pacific-Asia
  Conference on Knowledge Discovery and Data Mining}.\hskip 1em plus 0.5em
  minus 0.4em\relax Springer, 2010, pp. 362--373.

\bibitem{sayyadi2009futurerank}
H.~Sayyadi and L.~Getoor, ``Futurerank: Ranking scientific articles by
  predicting their future pagerank,'' in \emph{Proceedings of the 2009 SIAM
  International Conference on Data Mining}.\hskip 1em plus 0.5em minus
  0.4em\relax SIAM, 2009, pp. 533--544.

\bibitem{Dunbar1992}
R.~I.~M. Dunbar, ``Neocortex size as a constraint on group size in primates,''
  \emph{Journal of Human Evolution}, vol.~22, no.~6, pp. 469--493, 1992.

\end{thebibliography}
